\documentclass[fleqn,usenatbib]{mnras}

\usepackage{newtxtext,newtxmath}

\usepackage[T1]{fontenc}

\DeclareRobustCommand{\VAN}[3]{#2}
\let\VANthebibliography\thebibliography
\def\thebibliography{\DeclareRobustCommand{\VAN}[3]{##3}\VANthebibliography}

\usepackage{graphicx}	
\usepackage{amsmath}	
\usepackage{CJKutf8}

\newcommand{\teff}{T_\mathrm{eff}}
\newcommand{\logg}{\log{g}}
\newcommand{\loggf}{\log{gf}}
\newcommand{\feh}{\mathrm{[Fe/H]}}
\newcommand{\pfe}{\mathrm{[P/Fe]}}
\newcommand{\ap}{A(\mathrm{P})}
\newcommand{\ac}{A(\mathrm{C})}
\newcommand{\iniZ}{\mathrm{ini}Z}
\newcommand{\vmic}{V_\mathrm{mic}}
\newcommand{\vmac}{V_\mathrm{mac}}
\newcommand{\vbroad}{V_\mathrm{broad}}
\newcommand{\vsini}{v\sin{i}}

\newcommand{\CNnames}[1]{{\begin{CJK}{UTF8}{gbsn}~(#1)~\end{CJK}}}

\title[SPA P abundance]{Stellar population astrophysics (SPA) with the TNG. The Phosphorus abundance on the young side of Milky Way\thanks{Based on observations made with the Italian Telescopio Nazionale Galileo (TNG) operated on the island of La Palma by the Fundación Galileo Galilei of the INAF (Istituto Nazionale di Astrofisica) at the Spanish Observatorio del Roque de los Muchachos of the Instituto de Astrofisica de Canarias}}

\author[M. Jian et al.]{
Mingjie Jian\CNnames{简明杰},$^{1}$\thanks{E-mail: jian-mingjie@outlook.com}
Xiaoting Fu\CNnames{符晓婷},$^{2,3}$
Valentina D'Orazi,$^{4,5}$
Angela Bragaglia,$^{3}$
\newauthor
S. Bijavara Seshashayana,$^{6, 7}$
He Zhao\CNnames{赵赫},$^{8}$
Ziyi Guo\CNnames{郭子怡},$^{9,10}$
Karin Lind,$^{1}$
\newauthor
Noriyuki Matsunaga\CNnames{松永典之},$^{11}$
Antonino Nunnari,$^{12, 4}$
Giuseppe Bono,$^{4}$
Nicoletta Sanna,$^{13}$
\newauthor
Donatella Romano,$^{3}$
and Marina Dal Ponte$^{5}$
\\
$^{1}$Department of Astronomy, Stockholm University, AlbaNova University Center, Roslagstullsbacken 21, 114 21 Stockholm, Sweden\\
$^{2}$Purple Mountain Observatory, Chinese Academy of Sciences, Nanjing 210023, China\\
$^{3}$INAF – Osservatorio di Astrofisica e Scienza dello Spazio di Bologna, via P. Gobetti 93/3, 40129 Bologna, Italy\\
$^{4}$Department of Physics, University of Rome Tor Vergata, via della Ricerca Scientifica 1, 00133, Rome, Italy\\
$^{5}$INAF - Osservatorio Astronomico di Padova, Vicolo dell' Osservatorio 5, 35122 Padova, Italy
\\
$^{6}$Materials Science and Applied Mathematics, Malm\"o University, SE-205 06 Malm\"o, Sweden\\
$^{7}$Nordic Optical Telescope, Rambla Jos\'e Ana Fern\'andez P\'erez 7, ES-38711 Bre\~na Baja, Spain\\
$^{8}$Departamento de Ciencias Fisicas, Universidad Andres Bello, Republica 220, 8320000 Santiago, Chile\\
$^{9}$School of Astronomy and Space Science, Nanjing University, Nanjing 210093, China\\
$^{10}$Key Laboratory of Modern Astronomy and Astrophysics (Nanjing University), Ministry of Education, Nanjing 210093, China\\
$^{11}$ Department of Astronomy, School of Science, The University of Tokyo, 7-3-1 Hongo, Bunkyo-ku, Tokyo 113-0033, Japan\\
$^{12}$ INAF – Astronomic Observatory of Rome, Via Frascati 33, 00078 Monte Porzio Catone, Italy\\
$^{13}$ INAF – Osservatorio Astrofisico di Arcetri, Largo E. Fermi 5, 50125 Firenze, Italy\\
}

\date{Accepted XXX. Received YYY; in original form ZZZ}

\pubyear{\the\year{}}

\begin{document}
\label{firstpage}
\pagerange{\pageref{firstpage}--\pageref{lastpage}}
\maketitle

\begin{abstract}
We present phosphorus abundance measurements for a total of 102 giant stars, including 82 stars in 24 open clusters and 20 Cepheids, based on high-resolution near-infrared spectra obtained with GIANO-B.
Evolution of phosphorus abundance, despite its astrophysical and biological significance, remains poorly understood due to a scarcity of observational data.
By combining precise stellar parameters from the optical, a robust line selection and measurement method, we measure phosphorus abundances using available \ion{P}{i} lines.
Our analysis confirms a declining trend in $\mathrm{[P/Fe]}$ with increasing $\mathrm{[Fe/H]}$ around solar metallicity for clusters and Cepheids, consistent with previous studies.
We also report a $\mathrm{[P/Fe]}$--age relation among open clusters older than 1\,Gyr, indicating a time-dependent enrichment pattern.
Such pattern can be explained by the different stellar formation history of their parental gas, with more efficient stellar formation in the gas of older clusters (thus with higher phosphorus abundances).
$\mathrm{[P/Fe]}$ shows a flat trend among cepheids and clusters younger than 1\,Gyr (along with three Cepheids inside open clusters), possibly hinting at the phosphorus contribution from the previous-generation low-mass stars.
Such trend suggests that the young clusters share a nearly common chemical history, with a mild increase in phosphorus production by low-mass stars.

\end{abstract}




\section{Introduction}

Spectroscopic analysis of stars allows us to determine the abundances of various elements in their atmospheres, effectively revealing their chemical compositions.
These elemental signatures reflect the composition of the molecular cloud from which the stars were born, capturing the initial conditions of star formation.
Consequently, measuring the chemical compositions of stars across different elements provides crucial constraints on the evolutionary history of the Milky Way and constitutes one of the central goals of Galactic Archaeology \citep[or Galactic Paleontology; see e.g.][]{Tolstoy2011}.
For instance, observations of $\alpha$-elements (e.g., O, Mg, Si, S, Ca and Ti) in the Milky Way help distinguish different components of the Galactic disk (e.g., the thin and thick disk; \citealt{Hayden2015, Anders2017}).
When combined with our understanding of the nucleosynthetic origins of these elements (e.g., \citealt{Kobayashi2020}), such measurements offer key insights into the chemical evolution of our host galaxy, solar system or even life \citep{P_life}.

However, the chemical evolution of some elements remains elusive due to a lack of observational constraints.  
Phosphorus is one such element.
The observable spectral lines of phosphorus are mostly located in the ultraviolet and infrared wavelength regions, making it impossible to measure its abundance using optical spectra. 
Given that ultraviolet observations are heavily affected by interstellar extinction, the infrared becomes the most suitable wavelength range for studying phosphorus abundance and thus provides the most useful information for Galactic archaeology. 
In contrast, many spectral lines of other elements are found in the optical, which makes it easier to determine stellar parameters using optical spectra. 
These parameters can then be used to infer the abundances of elements whose lines are only available in the infrared. 
Simultaneously obtained optical and infrared spectra are particularly well-suited for this approach (see, e.g., \citealt{Jian2024}).


Observational studies of phosphorus began relatively recently, with the pioneering work of \citet{Caffau2011}, who measured the P abundances of 20 F-type dwarf stars.
They found that the $\mathrm{[P/Fe]}$ ratio decreases from approximately $+0.4$ to $0$ as metallicity increases from $\mathrm{[Fe/H]} = -1$ to solar.
This trend contrasts with that of other light odd-$Z$ elements, such as Na and Al, and also deviates from the theoretical predictions of \citet{Kobayashi2006}.
Subsequently, \citet{Roederer2014} extended the observational data towards the metal-poor regime.
Using the Hubble Space Telescope Imaging Spectrograph, they measured phosphorus abundances in 14 dwarf stars with metallicities ranging from $\mathrm{[Fe/H]} = -4$ to $-0.1$\,dex.
$\mathrm{[P/Fe]}$ remains approximately solar in the metallicity range $\mathrm{[Fe/H]} \sim -4$ to $-2$\,dex, then increases to $\mathrm{[P/Fe]} \sim 0.5$ at $\mathrm{[Fe/H]} \approx -1.1$, before decreasing back to near-solar levels at higher metallicities.
Most subsequent studies have confirmed this general trend.
\citet{Hawkins2016} performed the first large-scale measurement of phosphorus abundances using data from the APOGEE survey, significantly expanding the sample around solar metallicity.
Later \citet{Hayes2022} measured phosphorus abundances for more than 120\,000 stars using APOGEE DR17, though only the upper-limit can be estimated for the majority of the stars ($\sim$87\,000) thus the completeness in metallicity is largely affected.
Over 40 additional measurements were later contributed by \citet{Caffau2016} and \citet{Caffau2019}, using the high-resolution spectrograph GIANO.
\citet{Nandakumar2022} used the IGRINS spectrograph to reproduce the $\mathrm{[P/Fe]}$ trend within the metallicity range of $\mathrm{[Fe/H]} = -1.2$ to $+0.3$\,dex.
In addition, \citet{Maas2022} found that the thick disc stars with the metallicity ranges from $\mathrm{[Fe/H]} = -1$ to $-0.4$\,dex tend to exhibit higher phosphorus abundances.
Interestingly, \citet{Weinberg2019} reported that the $\mathrm{[P/Mg]}$ ratio varies among stars with different magnesium abundances.
Since Mg is almost a ``pure'' $\alpha$ element that is mainly produced by core-collapse supernovae (CCSN), the results of \citet{Weinberg2019} contradict the theoretical prediction that phosphorus is predominantly produced by CCSN.

A number of giant stars with peculiar phosphorus abundances have also been identified.
\citet{Masseron2020a} discovered 15 stars exhibiting extremely high phosphorus enhancements ($[\mathrm{P}/\mathrm{Fe}] \gtrsim 1.5$) with unusual overabundances O, Mg, Si, Al, and Ce, and further examined the detailed abundance patterns of three of these stars in a follow-up study \citep{Masseron2020b}.
This sample was later expanded to 78 stars by \citet{Brauner2023}.
The unusual abundance patterns observed in these objects continue to pose a challenge to current Galactic chemical evolution (GCE) models.

\begin{figure}
    \centering
    \includegraphics[width=1\linewidth]{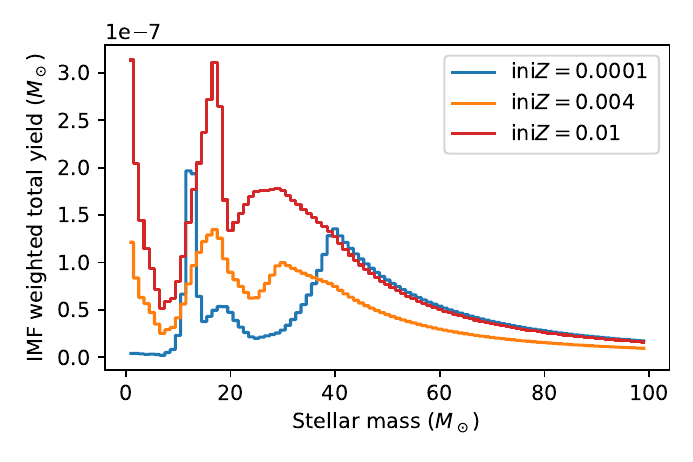}
    \caption{Stellar total yield of phosphorus in three different initial metallicities ($\iniZ$). The yield has been weighted by Kroupa IMF, and the stellar population has been normalized to 1 $\rm M_{\odot}$.
    }
    \label{fig:P-yield}
\end{figure}

GCE models suggest that phosphorus is primarily produced in massive stars.
Figure~\ref{fig:P-yield} presents the stellar total yield of phosphorus (the amount of element newly-produced plus that present in the star at birth, in contrast with the net yield, those newly-produced by stellar nucleosynthesis) after being weighted by the Kroupa stellar initial mass function \citep [IMF, ][]{Kroupa2001}.
We consider three types of yields here: low-mass stars (from \citealt{Karakas2010}), massive stars (from \citealt{Nomoto2013})  and Type~Ia supernovae (SNIa, from \citealt{Iwamoto1999}).
The massive stars contribute over 85\% of phosphorus in a single stellar population in all three metallicities we consider for a representative initial metallicities for Population~I and II progenitor stars in galactic chemical evolution models.
Also, the yield for metal-rich stars is higher than that for metal poor stars in most of the mass range, since in general stellar wind increases as metallicity increases \citep{Vink2001}.
\citet{Cescutti2012} compares their phosphorus measurement on stars with $\mathrm{[Fe/H]} > -1$  with the GCE models, and concluded that the major producer of P should be CCSN, though their adopted yields need to be multiplied by 3 to match the [P/Fe] values, and their trend does not match the metal-poor observation from \citet{Roederer2014}.
Recently \citet{Bekki2024} suggested that the P production of oxygen-neon novae (ONe novae) needs to be included to explain the solar [P/Fe] values in the range of [Fe/H] $< -2.5$ dex as well as the increase between $-2.5 < \feh < -1$.

To unfold the evolutionary history of phosphorus in the Milky Way, stellar age is a key parameter.
Despite recent observational progress, the age--phosphorus relation remains poorly understood.
Determining the ages of individual field stars in the Milky Way with high precision is challenging.
\citet{Maas2019} attempted to explore this relation by constructing a P--age diagram, but the stellar ages in their sample suffer from large uncertainties.
They emphasised that more accurate age determinations are essential to unveil the evolutionary behaviour of phosphorus in the thin disc of the Milky Way.

Cluster member stars, on the other hand, provide more reliable age estimates by fitting isochrones to the cluster colour--magnitude diagram (CMD).
This offers a promising avenue for constructing a more precise P--age relation.
In this work, we present phosphorus abundance measurements for members of 24 open clusters as well as 20 Cepheids in the field, located in different directions across the Galaxy and the Galactocentric distance ($R_\mathrm{gc}$) between $7$ to $10\,$kpc, aiming to place new constraints on the chemical evolution of phosphorus in the Milky Way.

This paper is organised as follows.
Section~\ref{sec:data} describes the data and the reduction process.
Section~\ref{sec:P-mesurement} outlines the method used to determine phosphorus abundances.
The results are presented in Section~\ref{sec:result}, followed by a discussion in Section~\ref{sec:discussion} and conclusions in Section~\ref{sec:conclusion}.




\section{Data and reduction}
\label{sec:data}

The data used in this study consist of near-infrared (NIR) spectra covering the wavelength range $0.9$--$2.45\,\mu\mathrm{m}$, obtained with a spectral resolution of $R = 50{,}000$ using the GIANO-B spectrograph \citep{Oliva2012a, Oliva2012b, Origlia2014}.
These NIR spectra were acquired simultaneously with optical spectra spanning $0.383$--$0.690\,\mu\mathrm{m}$ at a resolution of $R = 115{,}000$, obtained with the HARPS-N spectrograph \citep{Cosentino2012} in GIARPS mode \citep{Tozzi2016, Claudi2017} at the Telescopio Nazionale Galileo (TNG), a 3.58-m optical/infrared telescope located in the 
Roque de los Muchachos Observatory in La Palma, Canary Islands.
This configuration effectively eliminates temporal variations between the NIR and optical spectral lines.
The dataset is part of the Large Programme Stellar Population Astrophysics (SPA), which aims to derive detailed, age-resolved chemical abundances across the Milky Way disc (programme ID A37TAC\_31, PI: L. Origlia).
The programme began in 2018 and was awarded observing time with both the HARPS-N and GIANO-B high-resolution echelle spectrographs at the TNG.

\begin{figure*}
    \centering
    \includegraphics[width=0.95\linewidth]{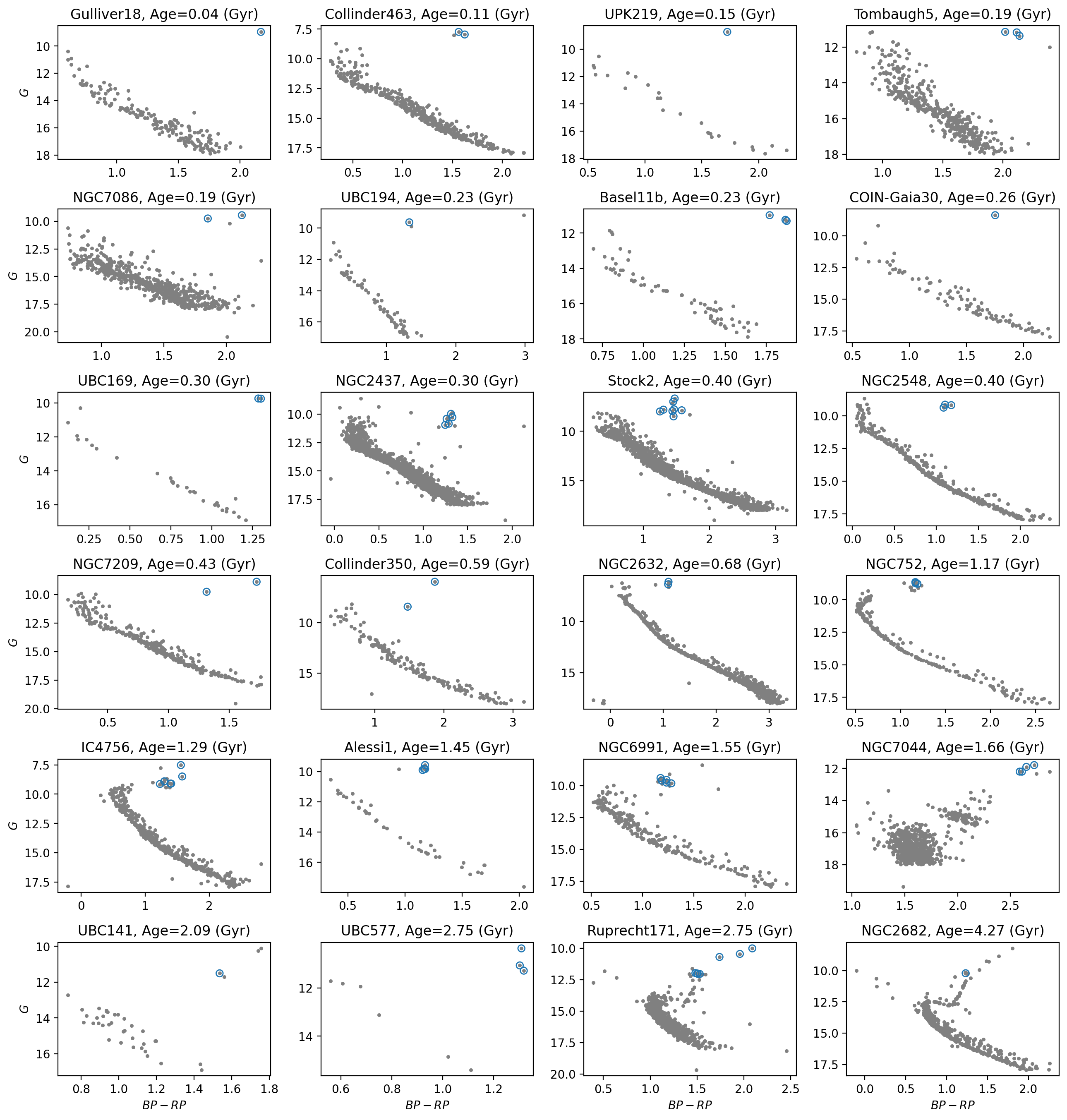}
    \caption{CMD of the target clusters. The grey dots represent stars with membership probabilities greater than 0.5 from \citet{Cantat-Gaudin2020}, while the blue circles highlight our selected cluster giant stars.}
    \label{fig:cluster-cmd}
\end{figure*}

Our dataset consists of 82 giant stars located in open clusters, supplemented by 20 field Cepheids and two standard stars: the Sun and Arcturus.
Table~\ref{tab:obslog} provides the observational log of the target stars, and Figure~\ref{fig:cluster-cmd} displays the CMDs of our sample.
We adopted the ages of the clusters from \citet{Cantat-Gaudin2020} in this study.
For the Cepheids, we derived the ages from their period using the $Z=0.02$ period-age relation presented in \citet{Bono2005}, with their pulsation mode (fundamental or first overtone) distinguished.
Three of the Cepheids, DL Cas, SV Vul and X Vul, are classified as open cluster members (e.g., in \citealt{Hao2022}).
SV Vul and X Vul have similar ages from their pulsation period and host clusters (with differences $<12\,$Myr).
DL Cas, however, has a age from period-age relation of $51\pm9\,$Myr, while its host cluster, NGC\,129 is $\sim 130\,$Myr old.
Considering that only three of our Cepheids have ages from their host clusters, we adopt their ages from period-age relation to increase the number of stars with age measurement.

Although the SPA project also includes many dwarf stars in these clusters, measuring phosphorus abundances for them is particularly challenging.
This is because, in dwarfs, phosphorus lines are significantly weaker than in other types of stars (see Figure~\ref{fig:cluster-cmd}), while their typically higher $\vsini$ values broaden the lines. This combination highly reduce the P line detectability in dwarfs.
In addition, dwarfs are intrinsically fainter, and the P lines are often smeared out by noise in their spectra.
Thus we limit our target stars to giants in this study.

\begin{table*}
    \centering
    \caption{Observation log of our target stars, with the columns of (left to right) star name, Gaia DR3 ID, observation date in reduced heliocentric Julian dates ($\mathrm{RHJD} = \mathrm{HJD} - 2400000$), exposure time, and the mean signal-to-noise ratio per pixel. Only the first six stars from cluster giant and Cepheid sample and the standard stars are listed, and the full table is available in CDS.}
    \begin{tabular}{lllll}
    \hline
    Star name & Gaia DR3 ID & RHJD & $t_\mathrm{exp}$\,(s) & $S/N_\mathrm{mean}$\\
    \hline\hline
    Alessi1-2 & 402506369136008832 & $58471.820$ & $300\times12$ & $212$ \\
    Alessi1-3 & 402505991180022528 & $58471.872$ & $300\times12$ & $225$ \\
    Alessi1-5 & 402867593065772288 & $58470.864$ & $300\times8$ & $233$ \\
    Alessi1-6 & 402880684126058880 & $58471.926$ & $300\times10$ & $199$ \\
    Alessi Teutsch11-1 & 2184332753719499904 & $58710.863$ & $300\times2$ & $377$ \\
    Basel11b-1 & 3424056131485038592 & $58513.903$ & $300\times10$ & $196$ \\
    ... & ... &... &... & ... \\
    \hline
    DL Cas (NGC 129) & 428620663657823232 & $58354.228$ & $300\times4$ & $253$ \\
    SV Vul (UBC 130) & 2027951173435143680 & $58425.869$ & $300\times2$ & $545$ \\
    X Vul (UBC 129) & 2027263738133623168 & $58349.853$ & $300\times4$ & $375$ \\
    CO Aur & 3451987987438434944 & $58426.248$ & $300\times4$ & $398$ \\
    RX Aur & 200708636406382720 & $58429.227$ & $300\times4$ & $425$ \\
    RW Cam & 473043922712140928 & $58354.187$ & $300\times4$ & $388$ \\
    ... & ... &... &... & ... \\
    \hline
    Sun (Ganymede) & -- & $58325.880$ & $30\times2$ & $130$\\
    Arcturus & -- & $58301.894$ & $60\times2$ & $612$ \\
    \hline
    \end{tabular}
    \label{tab:obslog}
\end{table*}

\subsection{Telluric correction and normalization}

The GIANO-B spectra were reduced using the GOFIO data reduction pipeline \citep{Rainer2018}.
For each spectral order, the pipeline performs bad pixel and cosmic ray removal, sky and dark subtraction, and corrections for flat-field and blaze effects, followed by the extraction of one-dimensional spectra.
We further processed the extracted spectra using the pre-analysis pipeline \texttt{giano\_ct}\footnote{\url{https://github.com/MingjieJian/giano_ct}}, which carries out continuum normalisation and telluric absorption correction.
Continuum normalisation is performed using the ``alpha-roll'' method described by \citet{Xu2019} and \citet{Cretignier2020}.
In this method, a circle with radius $\alpha$ is rolled along the top of the spectrum, and the contact points between the circle and the flux curve are selected as continuum points.
The value of $\alpha$ is adaptively determined based on the difference between the smoothed and observed spectra, increasing in regions where absorption lines are present.
Telluric correction is then applied using the \texttt{TelFit} package \citep{Gullikson2014}, by fitting the model telluric spectrum to the observed spectrum, and the individual spectral orders are merged into a single, continuum-normalised spectrum.
The best-fit telluric model is also saved for later analysis.
We note that certain wavelength regions ($13530$--$14350$\,\AA{} and $18020$--$19400$\,\AA{}) are heavily affected by telluric absorption and are therefore excluded during the correction process.

\section{Measurement of P abundance}
\label{sec:P-mesurement}

\subsection{Stellar parameters}
\label{sec:stellar-paras}

Stellar parameters (i.e., effective temperature $T_{\mathrm{eff}}$, surface gravity $\logg$, iron abundance [Fe/H] as a proxy of metallicity, and microturbulance velocity $\vmic$) for our target stars were independently determined by \citet{DalPonte2025} based on HARPS-N spectra, and by Bijavara-Seshashayana et al. (in prep.) using GIANO-B spectra.  
The two sets of parameters are generally consistent, although some discrepancies in $\logg$ are present for individual stars.  
In this work, we adopt the stellar parameters from \citet{DalPonte2025}.

The spectral synthesis in the following analysis is performed using the \texttt{PySME} code \citep{Wehrhahn2023}.
Spectral synthesis is carried out by the \texttt{SME} library \citep{Piskunov2017, Valenti1996}, based on a one-dimensional stellar atmosphere model, under the assumption of local thermodynamic equilibrium (LTE).
We note that although non-LTE corrections are available for 16 elements \citep{Amarsi2020, Mallinson2024}, there is currently no grid of departure coefficients for phosphorus.
\texttt{PySME} is capable of synthesising model spectra and determining stellar parameters and abundances by $\chi^2$ minimisation between the observed and synthetic spectra within user-defined masks.
The current version of \texttt{PySME} takes all input line lists into account during synthesis, which can be time- and memory-consuming when dealing with spectra that span a wide wavelength range.
Our recent developments in \texttt{PySME} include a more efficient synthesis via the dynamic removal of weak spectral lines, measurement of stellar parameters using pre-defined masks, and abundance measurements for multiple elements, all performed in an automated manner.
These improvements will serve as part of the pipeline for future large-scale spectroscopic surveys such as 4MOST \citep{deJong2019}.
The first two developments will be discussed and tested in detail in Jian et al.\ (in prep.), while we focus on the abundance measurement aspect in this article.

The line list, covering the wavelength range from $9360$ to $24,240$\,\AA, is extracted from the VALD3 database version 3750M \citep{Piskunov1995, Ryabchikova2015}.
We used the ``Extract All'' mode with hyperfine structure included.
All molecular lines across the full GIANO wavelength range are included, except for TiO lines.
The enormous number of TiO lines makes it nearly impossible both to download and to include them in the synthesis.
However, for target stars with $\teff \lessapprox 4500\,$K (14 stars in our sample), the blending of TiO lines may not be negligible.
Thus, we included the TiO lines within $\pm 20$\,\AA{} of the P lines used for abundance measurements for our target giants (i.e., those in Table~\ref{tab:p_lines_all} with $N_\mathrm{giant}$ not marked as `-').
We adopt MARCS model atmospheres \citep{Gustafsson2008} as input, and solar abundances are set according to \citet[][except for phosphorus; see Section~\ref{sec:sun_arcturus_result}]{Grevesse2007}.

\begin{table}
    \centering
    \caption{List of \ion{P}{i} lines used in this study, including their air wavelengths ($\lambda_\mathrm{air}$), lower excitation potentials (EP), and oscillator strengths ($\log gf$). Columns $N_\mathrm{giant}$ and $N_\mathrm{Cepheid}$ indicates the number of stars using the corresponding line for P measurement (excluding cases where the line is blended with telluric features or only provides an upper limit), and the number in the parentheses (if presents) indicates the number of stars being removed (see the description in section~\ref{sec:abund_measure}). The final column (`Sun') marks whether the line was also used in the solar abundance determination.
    }
    \begin{tabular}{cccccc}
    \hline
    $\lambda_\mathrm{air}$ & EP & $\loggf$ & $N_\mathrm{giant}$ & $N_\mathrm{Cepheid}$ & Sun\\
    ({\AA}) & (eV) &  & & &\\
    \hline\hline
        $9609.036$ & $6.9356$ & $-1.05$ & - & $0$ ($1$) &  \\
        $9676.222$ & $8.0785$ & $0.00$ & - & $4$ &  \\
        $9734.755$ & $6.9543$ & $-0.36$ & - & $3$ &  \\
        $9750.748$ & $6.9543$ & $-0.18$ & $12$ & $1$ & Y \\
        $9790.194$ & $7.1758$ & $-0.69$ & - & $2$ &  \\
        $9796.828$ & $6.9852$ & $0.27$ & - & $1$ & Y \\
        $9903.671$ & $7.1758$ & $-0.30$ & - & $7$ & Y \\
        $9976.681$ & $6.9852$ & $-0.29$ & - & $1$ ($5$) & Y \\
        $10084.277$ & $7.2127$ & $0.14$ & - & $4$ &  \\
        $10204.716$ & $7.2127$ & $-0.52$ & - & $4$ &  \\
        $10511.588$ & $6.9356$ & $-0.13$ & $11$ & $1$ &  \\
        $10529.524$ & $6.9543$ & $0.24$ & $22$ & $1$ & Y \\
        $10581.577$ & $6.9852$ & $0.45$ & $34$ & $1$ & Y \\
        $10596.903$ & $6.9356$ & $-0.21$ & $29$ ($1$) & $4$ & Y \\
        $10681.406$ & $6.9543$ & $-0.19$ & $6$ & $5$ & Y \\
        $10769.511$ & $6.9543$ & $-1.07$ & - & $2$ &  \\
        $10813.141$ & $6.9852$ & $-0.41$ & $0$ ($12$) & $5$ &  \\
        $10932.724$ & $8.0785$ & $0.31$ & - & $1$ &  \\
        $10967.373$ & $8.0783$ & $0.12$ & - & $7$ &  \\
        $11183.240$ & $7.2127$ & $0.40$ & $8$ & - &  \\
        $16254.749$ & $8.2255$ & $0.00$ & - & $7$ &  \\
        $16482.932$ & $7.2127$ & $-0.29$ & $42$ & $5$ & Y \\
        $16590.045$ & $8.2276$ & $0.50$ & - & $0$ ($6$) &  \\
        $16738.681$ & $8.2864$ & $0.00$ & - & $0$ ($4$) &  \\
        $17112.447$ & $8.2504$ & $0.50$ & $0$ ($51$) & $0$ ($7$) &  \\
        $17286.920$ & $8.2864$ & $0.00$ & - & $3$ &  \\
        $17423.670$ & $8.2864$ & $0.00$ & $0$ ($2$) & $4$ &  \\
    \hline
    \end{tabular}
    \label{tab:p_lines_all}
\end{table}

The broadening velocity, including macroturbulence and projected rotational velocity, $\vbroad$, is not provided in \citet{DalPonte2025}.
We therefore measure $\vbroad$ using the GIANO-B spectra.
For each star, we first fix the other stellar parameters and select isolated Fe\,I lines based on the synthetic spectrum (the details of the line selection method are described in Section~\ref{sec:line-selection}).
The $\vbroad$ value is then fitted for each selected Fe\,I line, and the weighted mean and standard deviation across all lines are adopted as the final broadening velocity and its associated uncertainty.

\begin{table*}
    \centering
    \caption{Stellar parameters of the target stars. Only the first six stars from cluster giant and Cepheid sample and the standard stars are listed, and the full table is available in CDS.}
    \begin{tabular}{cccccc}
    \hline
    Star name & $\teff$ & $\logg$ & [Fe/H] & $\vmic$ & $\vbroad$\\
    \hline\hline
    Alessi1-2 & $4986\pm30$ & $2.83\pm0.07$ & $-0.001\pm0.010$ & $1.297\pm0.013$ & $6.7\pm0.9$ \\
    Alessi1-3 & $4996\pm30$ & $2.84\pm0.07$ & $-0.007\pm0.010$ & $1.276\pm0.014$ & $6.6\pm0.9$ \\
    Alessi1-5 & $4939\pm31$ & $2.73\pm0.08$ & $-0.028\pm0.010$ & $1.353\pm0.013$ & $6.7\pm0.9$ \\
    Alessi1-6 & $4985\pm32$ & $2.87\pm0.08$ & $-0.013\pm0.010$ & $1.278\pm0.014$ & $6.4\pm0.8$ \\
    Alessi Teutsch 11-1 & $4517\pm35$ & $2.15\pm0.11$ & $-0.040\pm0.013$ & $1.756\pm0.014$ & $7.2\pm1.1$ \\
    Basel 11b-1 & $4950\pm35$ & $2.51\pm0.09$ & $0.007\pm0.010$ & $1.808\pm0.010$ & $8.4\pm1.1$ \\
    ... & ... & ... & ... & ... & ... \\
    \hline
    DL Cas & $5622\pm18$ & $1.73\pm0.04$ & $0.050\pm0.010$ & $3.920\pm0.040$ & $25.92\pm0.17$ \\
    SV Vul & $5676\pm13$ & $1.00\pm0.03$ & $0.190\pm0.010$ & $4.240\pm0.020$ & $15.19\pm0.06$ \\
    X Vul & $6294\pm12$ & $1.95\pm0.02$ & $0.170\pm0.010$ & $4.410\pm0.020$ & $18.04\pm0.06$ \\
    CO Aur & $7076\pm9$ & $2.77\pm0.01$ & $0.180\pm0.010$ & $2.670\pm0.020$ & $13.52\pm0.05$ \\
    RX Aur & $6466\pm10$ & $1.52\pm0.01$ & $-0.040\pm0.010$ & $4.040\pm0.020$ & $26.62\pm0.09$ \\
    RW Cam & $5672\pm16$ & $1.63\pm0.04$ & $0.170\pm0.010$ & $3.180\pm0.020$ & $17.51\pm0.09$ \\
    ... & ... & ... & ... & ... & ... \\
    \hline
    Sun & $5772$ & $4.44$ & $0.00$ & $1.0$ & $2.8$ \\
    Arcturus & $4286$ & $1.6$ & $-0.52$ & $1.74$ & $4.6$ \\
    \hline
    \end{tabular}
    \label{tab:stellar-paras}
\end{table*}

\begin{figure}
    \centering
    \includegraphics[width=1\linewidth]{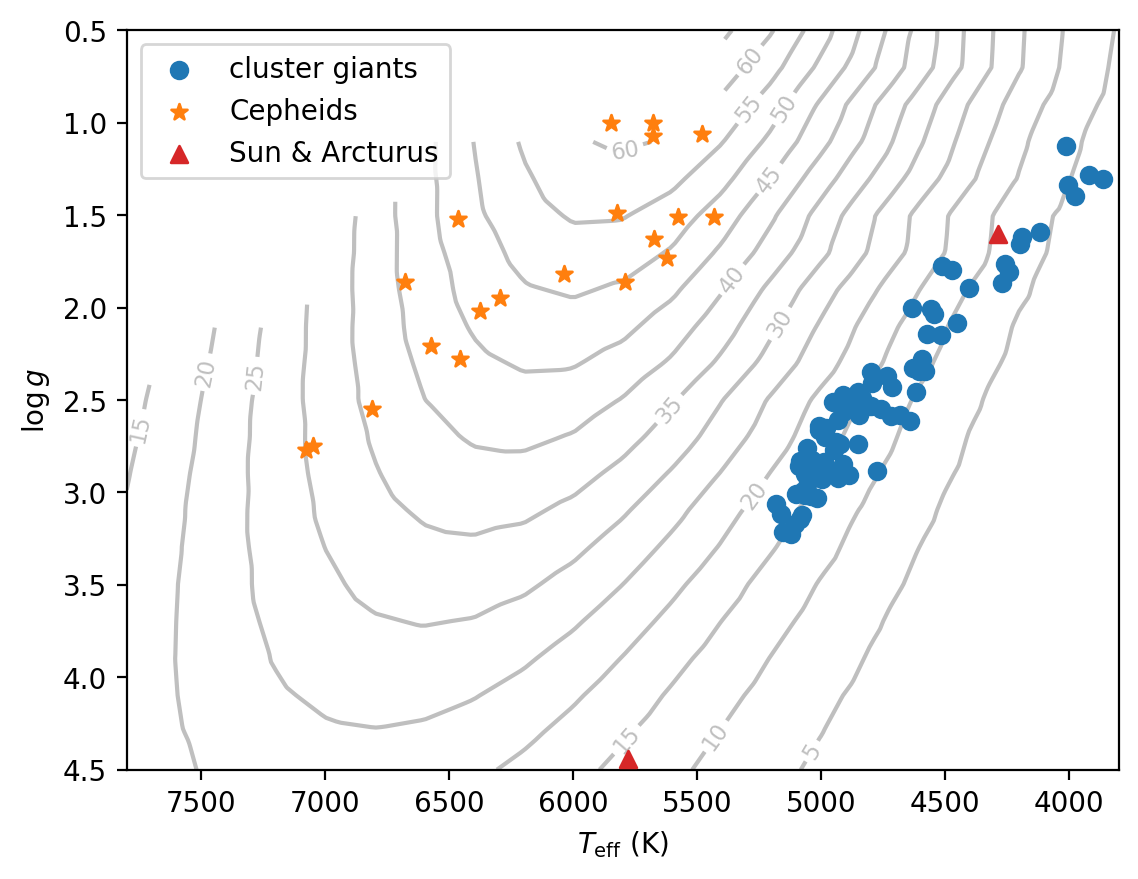}
    \caption{The Kiel diagram of our sample stars. The underlying contours show the EW (m\AA) of \ion{P}{i} line at 10529.524\,\AA{} at solar metallicity.}
    \label{fig:kiel-diagram}
\end{figure}

The adopted stellar parameters are listed in Table~\ref{tab:stellar-paras} and plotted in Figure~\ref{fig:kiel-diagram}.
The theoretical equivalent width (EW) of the \ion{P}{i} line at 10529.524\,\AA{} -- one of the strongest phosphorus lines in the near-infrared -- is also shown as a contour in the diagram.
Most infrared P lines exhibit EW contour shapes similar to this line, with maximum strengths occurring at approximately $T_{\mathrm{eff}} \approx 6000$\,K and low $\logg$, and decreasing as $T_{\mathrm{eff}}$ moves away from this peak or as $\logg$ increases.
As a result, most of our Cepheids exhibit relatively strong P lines (see also \citealt{Elgueta2024}).
In contrast, our cluster giants lie far from the EW peak region, and the EWs of the 10529.524\,\AA{} line are typically around $15$\,m\AA{}.
Therefore, high signal-to-noise ratio spectra are required to derive reliable phosphorus abundances for these stars.

\subsection{Line selection}
\label{sec:line-selection}

Given the relatively large sample size of stars in this study, as well as their wide range of stellar parameters, it is essential to select suitable \ion{P}{i} lines using the synthetic spectra.
The VALD database contains 229 \ion{P}{i} lines within the relevant wavelength range.
However, whether a given line can be used for phosphorus abundance determination depends on several factors: whether the line appears in the observed spectrum, the degree of blending with nearby lines, and whether the line can be reliably measured given the quality of the observed spectra.
Therefore, a careful line selection process is required prior to performing the abundance analysis.

The goal of the line selection process is to identify all P lines that are present in the spectra and exhibit minimal blending, based on synthetic spectra.
All \ion{P}{i} lines are first extracted from the line list.
For each star, we define a total broadening velocity, $\vbroad$, which accounts for the broadening of microturbulence velocity ($\vmic$), macroturbulence velocity ($\vmac$), projected rotation ($\vsini$), and instrumental resolution ($R$), as:
\begin{equation}
    \vbroad = \sqrt{\vmic^2 + \vmac^2 + (\vsini)^2 + \left( \frac{c}{R} \right)^2},
\end{equation}
with $c$ as the speed of light.
A wavelength region with a width of $4\vbroad/c$ is assigned to each target line, defining a feature.
If two features overlap in wavelength, they are merged and treated as a single feature.

Synthetic spectra based on the stellar parameters mentioned in Section~\ref{sec:stellar-paras} are then generated for each feature, and three variables are used to evaluate the suitability of each P line for abundance analysis:
\begin{itemize}
    \item \textbf{Feature-dominance}: the ratio of the EW of the phosphorus line(s) to the total EW of all lines within the feature;
    \item \textbf{Feature-depth}: the maximum depth of the feature in the normalised flux;
    \item \textbf{Telluric-depth}: the maximum depth of telluric absorption in the corresponding wavelength region.
\end{itemize}

We require that the selected features satisfy the following thresholds: feature-dominance greater than 0.4, feature-depth greater than 0.006, and telluric-depth less than 0.1.
The adopted feature-dominance threshold does not require the feature to be completely dominated by phosphorus lines.
We found that the primary contaminants in P features are CN (at a wavelength range of $\lessapprox 12000$\,\AA) and CO ($\gtrapprox 12000$\,\AA) molecular lines.
To minimize the blending of these molecular lines, we altered the C abundance to fit the spectra of $\pm 2\,$\AA~around the P line and fix it during the fit for the P abundance.
These C abundances are presented in appendix~\ref{app:C-blend} and the tables therein.
The feature-depth threshold is determined based on the default synthesis precision of \texttt{PySME}.
The telluric-depth threshold of 0.1 is relatively strict.
Given that P lines are intrinsically weak, the profiles of P lines blended with deep telluric absorption are often significantly distorted.

A total of 11 and 26 \ion{P}{i} lines passed the automatic selection criteria for cluster giants and Cepheids, respectively, but
no \ion{P}{ii} line passed our selection.
Table~\ref{tab:p_lines_all} lists the line parameters and indicates which group of stars each line is used for.
The phosphorus lines that meet the threshold requirements for each target are subsequently included to derive its phosphorus abundance.
Minor refinements to the line selection are made following individual abundance measurements for each line.
The oscillator strength, or $\loggf$ value, is a key parameter that determines the strength of a spectral line in synthetic spectra.
Accurate abundance measurements therefore rely critically on the precision of the adopted $\loggf$ values.
While \citet{Elgueta2024} recalibrated the $\loggf$ values for seven of the P lines included in our selection, all of the selected lines already have laboratory-measured $\loggf$ values.
To maintain consistency across all line parameters, we adopt the $\loggf$ values provided by the VALD database for all P lines used in this work.

\subsection{Abundance measurement}
\label{sec:abund_measure}

\begin{figure*}
    \centering
    \includegraphics[width=1\linewidth]{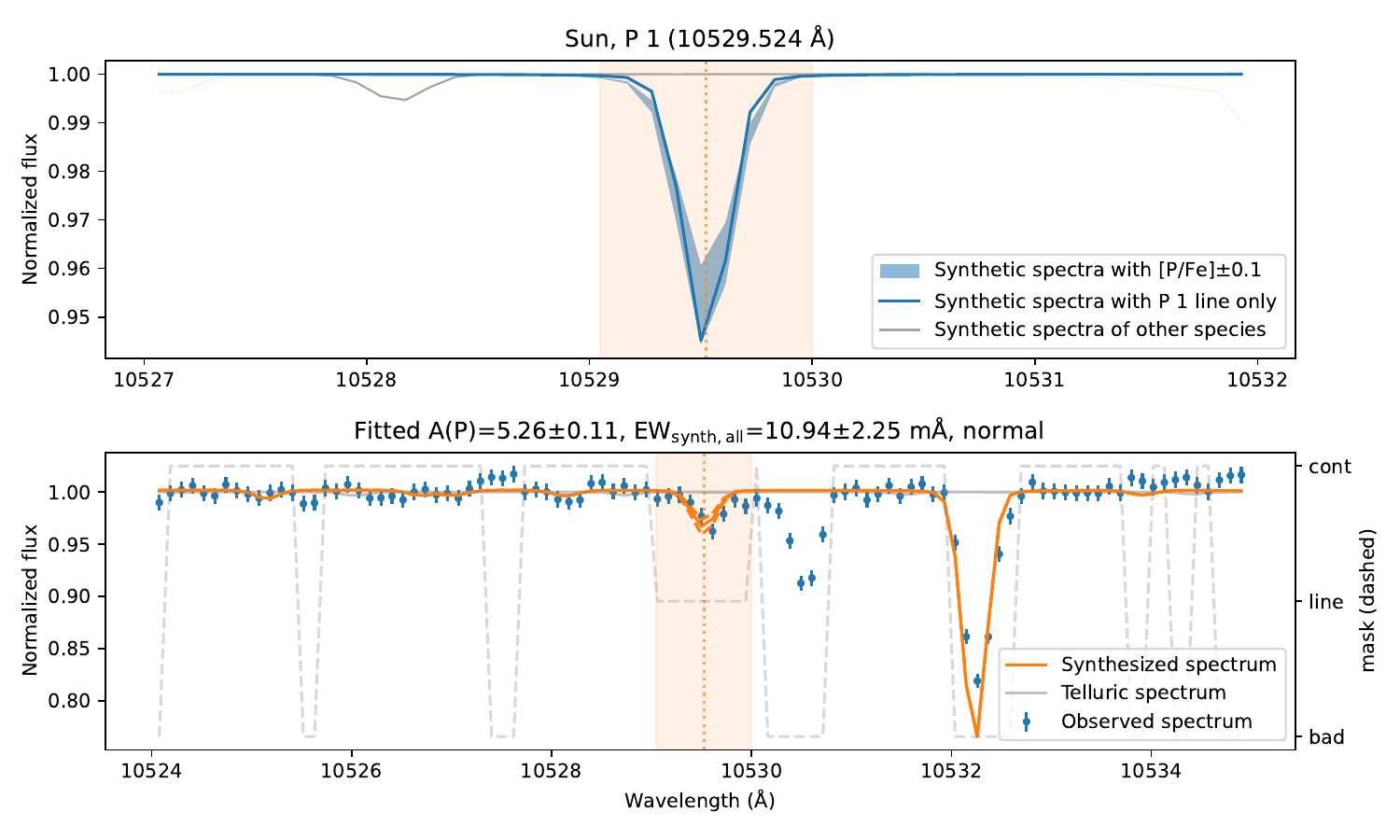}
    \caption{
    Phosphorus abundance determination from the \ion{P}{i} $10529.524\,$\AA{} line in the solar spectrum.
    \textbf{Top panel:} Synthetic spectra after varying the abundances of P by $\pm0.1$,dex (blue shaded profiles) together with spectra that contains only the \ion{P}{i} transition (solid blue), and other species (gray). The orange dotted vertical line marks the line centre, while the orange shaded strip marks the wavelength range used in the fit.
    \textbf{Bottom panel:} Observed spectrum (blue points) compared with the best–fitting synthetic spectrum (solid orange) and models corresponding to $\pm1\sigma$ uncertainty in $A(\mathrm{P})$ (orange dashed). The light-grey dashed ``stair-step'' curve (right-hand $y$-axis) shows the pixel mask: \textit{cont} (continuum windows), \textit{line} (pixels included in the fit), and \textit{bad} (pixels excluded). The orange shaded region is the same fitting window as in the top panel. Similar plots for the other analysed P lines are available in the Supporting Information.}
    \label{fig:P-line-fit-exmaple-sun}
\end{figure*}

\begin{figure*}
    \centering
    \includegraphics[width=1\linewidth]{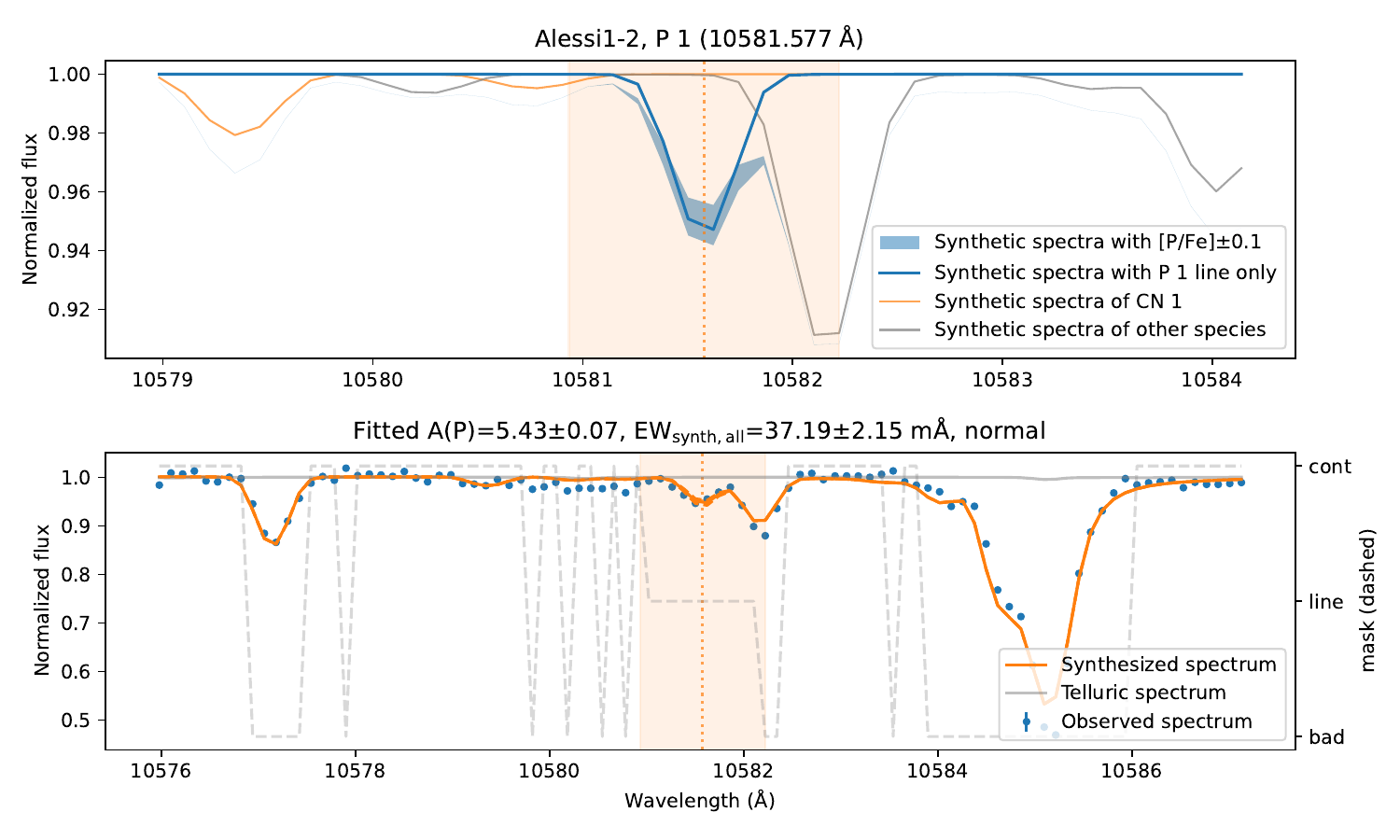}
    \caption{Similar as Figure~\ref{fig:P-line-fit-exmaple-sun}, but for the line at $10581.577\,$\AA{} of Alessi 1-2. Here several CN lines appear around the target P line, and their synthetic spectrum is also plotted in the upper panel. Similar plots for other lines and other stars are available in the Supporting Information.}
    \label{fig:P-line-fit-exmaple-Alessi_1_2}
\end{figure*}

For each phosphorus line, the pixels within the full feature region—defined as a width of $4\vbroad/c$ centred on the line—are used to fit the phosphorus abundance with \texttt{PySME}.
Figures~\ref{fig:P-line-fit-exmaple-sun} and \ref{fig:P-line-fit-exmaple-Alessi_1_2} show an example of the line fitting procedure, for our standard star the Sun and Alessi~1-2, respectively.
Synthetic spectra spanning a broader width of $4\vbroad/c + 4$\,\AA{} are generated, as illustrated in the upper panel.
Within the fitting region (vertical orange shaded band), if the blue curve (representing synthetic spectra generated using the P line only) lies between the shaded blue area (synthetic spectra with all lines, with varied [P/Fe]) and the blending by other species (the spectra with thin solid lines) is small, the feature is considered to be dominated by the P line.

The lower panel displays the configuration and result of the actual fitting.
To obtain a more accurate continuum level, we extend the input observed spectra to a total width of $4\vbroad/c + 10$\,\AA{}, centred on the P line.
Pixels are classified into three types of masks:
\begin{itemize}
    \item \textbf{Continuum mask}: pixels that satisfy all of the following conditions: (1) their depth in the synthetic spectrum is less than 0.025; (2) their depth in the telluric model is less than 0.1; and (3) the absolute difference between the observed and synthetic fluxes is less than twice the standard deviation of the residuals;
    \item \textbf{Line mask}: pixels within $4\vbroad/c$ of the P line centre;
    \item \textbf{Bad mask}: all remaining pixels.
\end{itemize}
This masking strategy ensures that pixels affected by other spectral lines or strong telluric absorption are excluded from continuum estimation.
As the continuum normalisation is generally well performed in Section~\ref{sec:data}, \texttt{PySME} applies only a constant scaling to the synthetic spectrum, determined by the mean flux level of the continuum pixels.
The phosphorus abundance is then derived by minimising the $\chi^2$ over the line mask region only.

We adopt the best-fit phosphorus abundance and the corresponding \texttt{fit\_uncertainties} (the statistical uncertainties of the fitting) output from \texttt{PySME} as our final measurement and its associated uncertainty.
Upper limits are identified following the method similar to the one described in \citet{Wang2024}:
if the EW of a fitted line at its best-fit abundance $A(\mathrm{P})_\lambda$ is smaller than three times the EW obtained at $A(\mathrm{P})_\lambda + \sigma_{A(\mathrm{P})_\lambda}$, the measurement is classified as an upper limit and excluded from further analysis.
When measurement uncertainties properly reflect the noise level in the spectrum, this criterion helps mitigate selection bias by avoiding the preferential inclusion of stronger lines influenced by noise.

We expect the phosphorus abundances derived from all selected lines for a given star to be consistent within their respective uncertainties.
However, we find that for a few lines, the derived abundances systematically deviate from the mean value obtained from the other lines.
This discrepancy may arise from several factors, such as inaccurate $\loggf$ values, improper continuum placement, or undetected blending features that are not captured by the synthetic spectra.
These problematic lines are marked with $N_\mathrm{giant}$ or $N_\mathrm{Cepheid}$ as 0 following by parentheses in Table~\ref{tab:p_lines_all} and are excluded from subsequent analysis.
Most of these lines are used in only a few stars, and we defer discussion of specific cases where the affected lines are present in the majority of stars to Appendix~\ref{app:unused-P}.

Finally, the weighted average and standard deviation of the $\ap$ from all available \ion{P}{i} lines are taken as the final $\ap$ and its corresponding error.

\section{Results}
\label{sec:result}

We present the results of phosphorus abundance measurements for individual stars and clusters in this section.

\subsection{Phosphorus Abundance for the standard stars}
\label{sec:sun_arcturus_result}

\begin{figure*}
    \centering
    \includegraphics[width=1\linewidth]{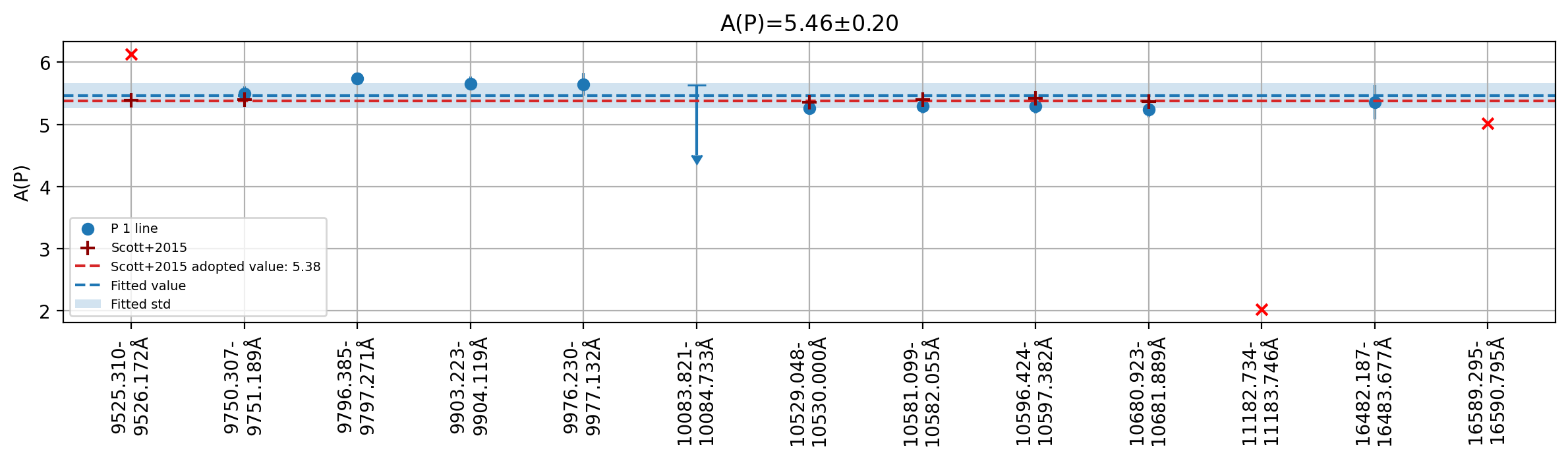}
    \caption{The Phosphorus abundance measurement result from all the lines for the Sun.
    Blue filled circles give the abundance $A(\mathrm{P})$ derived for each line (wavelength range shown on the $x$-axis); their vertical blue error bars mark the $1\sigma$ fitting uncertainty.
    Deep-red plus symbols plot the line-by-line abundances reported by \citet{Scott2015} for those that are used in both studies, enabling a direct comparison.
    Red crosses indicate lines discarded from our final mean because of severe telluric blending or being manually excluded, while blue downward arrows denote lines for which only an upper limit on $A(\mathrm{P})$ could be set.
    The horizontal blue dashed line is the weighted mean abundance from all accepted (``normal-flag’’) lines, with its $1\sigma$ dispersion shown by the light-blue band.
    For reference, the red dashed line marks the solar value $A(\mathrm{P})=5.38$ adopted by \citet{Scott2015}.
    The panel title lists the mean abundance and its standard deviation obtained in this work.
    }
    \label{fig:P-fit_sun}
\end{figure*}

Our measurement of the solar spectrum from GIANO shows good consistency with previous determinations, as shown in Figure~\ref{fig:P-fit_sun}.
The same line selection procedure as for other stars was applied to the standard stars.
A total of 13 phosphorus lines were initially selected.
Among them, one line is too weak to provide reliable measurements and is therefore treated as upper limits, while another three are significantly blended with telluric absorption features or not showing a clear absorption profile.
It is evident that the lines affected by telluric contamination tend to deviate from the remaining measurements (i.e., the $9525.741\,$\AA{} and $11183.24\,$\AA{} line in Figure~\ref{fig:P-fit_sun}), which justifies their exclusion from the final abundance calculation.
The final phosphorus abundance, $5.46$, is obtained as the weighted mean of the remaining reliable P lines.
This value is in good agreement with the solar phosphorus abundance derived using 1D LTE spectral synthesis in \citet{Scott2015}, $5.38$.
The estimated uncertainty of $0.2$\,dex likely arises from a combination of factors, including lower resolution and signal-to-noise ratio compared to the solar spectrum used in \citet{Scott2015}, uncertainties in the adopted $\loggf$ values, and imperfect continuum placement.

For Arcturus, no phosphorus lines pass our line selection criteria.
Compared to the Sun, Arcturus is cooler and has significantly lower surface gravity ($\logg$), which leads to weaker P lines (see Figure~\ref{fig:kiel-diagram}).
In addition, its relatively low metallicity ($\mathrm{[Fe/H]} = -0.51$) further reduces the line strengths, rendering all available phosphorus lines too weak for reliable abundance measurement.

Our results from the Sun and Arcturus help to define an approximate detection boundary for phosphorus abundance measurements at the resolution and signal-to-noise ratio (S/N) of our GIANO-B spectra.
Specifically, a metallicity of $\mathrm{[Fe/H]} \sim -0.5$ appears to be the lower limit for detecting P lines in cool giant stars, as demonstrated by the non-detection in Arcturus.
For warmer giants and dwarfs, our test using the solar stellar parameters suggests that only one single P line at $10529.524$\,\AA{} remains detectable at $\mathrm{[Fe/H]} = -1.0$, indicating a rough detection limit under such conditions.
This conclusion is consistent with previous high-resolution studies, particularly those by \citet{Caffau2011}, whose sample consists of dwarf stars with the most metal-poor star having a metallicity just above $\mathrm{[Fe/H]} = -1.0$.
To extend phosphorus measurements to lower metallicities, stars within the optimum $\teff$ and $\logg$ range (see Fig~\ref{fig:kiel-diagram}) should be observed at very high SNR, or observations should be done in ultraviolet (see \citealt{Roederer2014}).

\subsection{Phosphorus Abundance for our sample stars}
\label{sec:P-age}

\begin{figure*}
    \centering
    \includegraphics[width=1\linewidth]{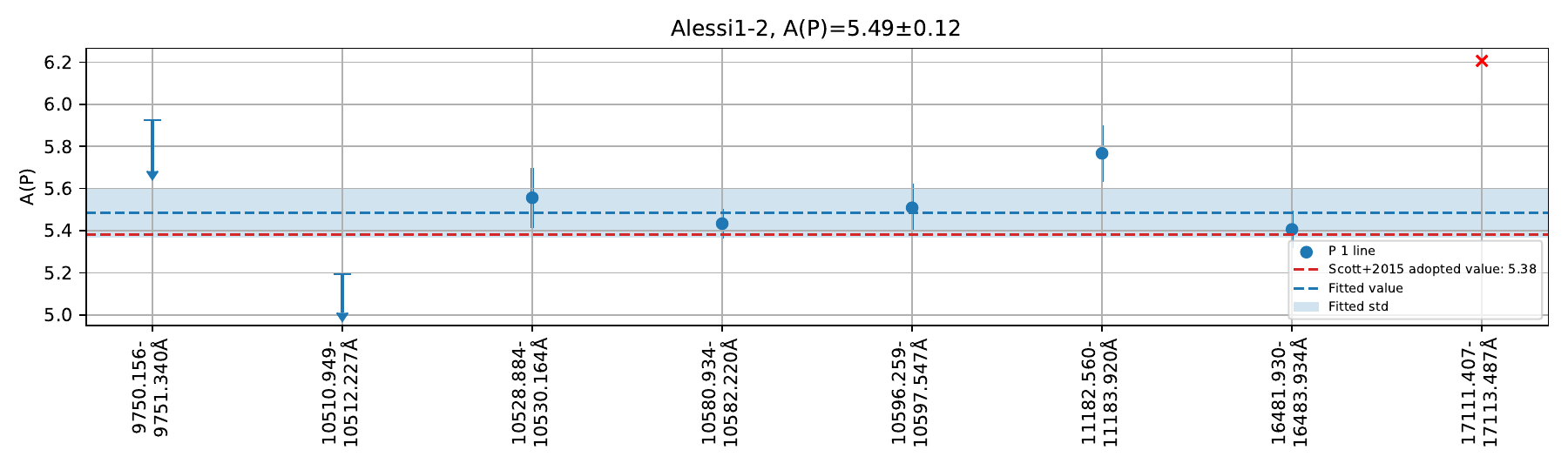}
    \caption{Similar as Figure~\ref{fig:P-fit_sun}, but for Alessi 2-1. Similar figures for other stars are available in the Supporting Information.
    }
    \label{fig:P-fit_example}
\end{figure*}

Figure~\ref{fig:P-fit_example} shows an example of the final phosphorus abundance determination for the star Alessi~1-2.
Eight lines were selected for the measurement, two of which exhibit weak absorption features and are therefore treated as upper limits.
One additional line was excluded based on the criteria described in Section~\ref{sec:abund_measure}.
The final phosphorus abundance is calculated as the weighted average of the abundances derived from the remaining lines.

Tables~\ref{tab:p-lines-giant} and \ref{tab:p-lines-cepheid} summarise the phosphorus abundance measurements for cluster giants and Cepheids, respectively.
The uncertainties in the averaged P abundances for the giants are typically around 0.1\,dex, while those for the Cepheids are mostly lower, indicating that P lines are generally stronger in Cepheid spectra.
The NLTE effect on the phosphorus lines may also contribute, at least in part, to the uncertainties in our target stars.
However, there is currently no available NLTE departure coefficient for phosphorus, and thus it is not yet possible to quantify or correct this effect.
We then compute the average phosphorus abundance for each cluster using its member stars, as listed in Table~\ref{tab:p-cluster}.

\begin{table*}
    \centering
    \caption{P measurement for cluster giants, with the star name, mean P abundance from all the normal-flag lines, number of normal-flag line and the abundance of the first six lines. The abundances with superscript of u or t are those with upper limit or blended by telluric lines. Only the first 10 stars and first 6 lines are listed, and the full table is available in CDS.}
    \begin{tabular}{clcllllll} 
    \hline
    star name & $\ap_\mathrm{mean}$ & $N$ & $\ap_{9750.748}$ & $\ap_{10511.588}$ & $\ap_{10529.524}$ & $\ap_{10581.577}$ & $\ap_{10596.903}$ & $\ap_{11183.24}$ \\
    \hline\hline
    Alessi 1-2 & $5.49\pm0.12$ & $5$ & $5.68\pm0.24^\mathrm{u}$ & $5.01\pm0.18^\mathrm{u}$ & $5.56\pm0.14$ & $5.43\pm0.07$ & $5.51\pm0.11$ & $5.77\pm0.13$ \\
    Alessi 1-3 & $5.50\pm0.14$ & $7$ & $5.80\pm0.12$ & $5.56\pm0.07$ & $5.20\pm0.09$ & $5.52\pm0.05$ & $5.55\pm0.12$ & $5.56\pm0.22$ \\
    Alessi 1-5 & $5.46\pm0.21$ & $5$ & $5.62\pm0.16$ & $4.98\pm0.18^\mathrm{u}$ & $5.37\pm0.09$ & - & $5.39\pm0.08$ & $5.72\pm0.09$ \\
    Alessi 1-6 & $5.51\pm0.23$ & $4$ & $5.62\pm0.25^\mathrm{u}$ & $5.16\pm0.19^\mathrm{u}$ & $5.28\pm0.18$ & - & $5.37\pm0.17$ & $5.84\pm0.15$ \\
    Alessi Teutsch 11-1 & $5.49\pm0.05$ & $2$ & $1.28\pm5.77^\mathrm{t}$ & $5.04\pm0.22^\mathrm{u}$ & $5.45\pm0.14$ & - & $5.49\pm0.06$ & - \\
    Basel 11b-1 & $5.41\pm0.13$ & $4$ & $4.76\pm0.82^\mathrm{u}$ & $4.95\pm0.31^\mathrm{u}$ & $5.45\pm0.17$ & - & $5.44\pm0.13$ & $5.53\pm0.16$ \\
    Basel 11b-2 & $5.35\pm0.19$ & $6$ & $5.92\pm0.14$ & $5.36\pm0.14$ & $5.27\pm0.22^\mathrm{u}$ & $5.30\pm0.04$ & $5.31\pm0.10$ & $5.73\pm0.16$ \\
    Basel 11b-3 & $5.46\pm0.14$ & $2$ & - & - & - & $5.51\pm0.09$ & - & - \\
    COIN-Gaia 30-1 & - & $0$ & $-1.36\pm37.53^\mathrm{t}$ & $-2.10\pm8.58$ & $5.42\pm0.28^\mathrm{u}$ & - & $4.70\pm0.41^\mathrm{r}$ & $5.82\pm0.11^\mathrm{t}$ \\
    Collinder 350-1 & - & $0$ & - & - & - & - & - & - \\
    Collinder 350-2 & $5.30\pm0.04$ & $2$ & - & - & - & $5.30\pm0.03$ & - & - \\
    ...             &...            &...  &...&...&...&...            &... &... \\
    \hline
    \end{tabular}
    \label{tab:p-lines-giant}
\end{table*}

\begin{table*}
    \centering
    \caption{Similar to Table~\ref{tab:p-lines-giant}, but for the sun and Cepheid sample, with ages coming from the period-age relation. Only the first 11 Cepheids are listed, and the full table is available in CDS.}
    \begin{tabular}{cccccccccc} 
    \hline
    star name & Age (Gyr) & $\ap_\mathrm{mean}$ & $N_\mathrm{line}$ & $\ap_{9734.755}$ & $\ap_{9750.748}$ & $\ap_{9790.194}$ & $\ap_{9796.828}$ & $\ap_{9903.671}$  & A(P)$_{9976.681}$ \\
    \hline\hline
    SV Vul & $0.016\pm0.003$ & $5.88\pm0.09$ & $1$ & $6.24\pm0.04^\mathrm{r}$ & - & - & - & - & - \\
    DL Cas & $0.051\pm0.009$ & $5.52\pm0.23$ & $4$ & - & - & $5.78\pm0.07$ & - & $5.61\pm0.11^\mathrm{t}$ & - \\
    CO Aur & $0.096\pm0.009$ & $5.73\pm0.17$ & $4$ & - & $4.63\pm0.93$ & - & - & - & - \\
    RX Aur & $0.039\pm0.007$ & $5.61\pm0.07$ & $4$ & - & - & $5.69\pm0.03$ & - & - & - \\
    V351 Cep & $0.080\pm0.008$ & $5.97\pm0.08$ & $2$ & - & $-1.76\pm6.86$ & - & - & - & - \\
    CD Cyg & $0.031\pm0.006$ & $5.65\pm0.08$ & $2$ & - & - & $5.70\pm0.21^\mathrm{t}$ & - & $5.75\pm0.11$ & - \\
    V1334 Cyg & $0.075\pm0.008$ & $5.61\pm0.13$ & $5$ & - & - & $5.74\pm0.10^\mathrm{t}$ & - & - & - \\
    VZ Cyg & $0.071\pm0.013$ & $5.65\pm0.25$ & $4$ & $6.19\pm0.22^\mathrm{t}$ & - & $5.44\pm0.45^\mathrm{t}$ & - & - & - \\
    X Cyg & $0.031\pm0.006$ & $5.43\pm0.07$ & $2$ & - & - & $5.63\pm0.13^\mathrm{t}$ & - & $5.98\pm0.08^\mathrm{t}$ & - \\
    W Gem & $0.051\pm0.009$ & $5.39\pm0.18$ & $7$ & - & - & - & - & - & - \\
    RR Lac & $0.059\pm0.011$ & $5.54\pm0.09$ & $5$ & - & - & - & $3.11\pm29.35^\mathrm{t}$ & - & - \\
    ... &... &... &... &... &... &... &... &... & ...\\
    \hline
    Sun & - & $5.46\pm0.20$ & $9$ & - & - & - & $5.50\pm0.12$ & - & $5.74\pm0.09$ \\
    \hline
    \end{tabular}
    \label{tab:p-lines-cepheid}
\end{table*}

\begin{table*}
    \centering
    \caption{The mean Phosphorus abundance of our target clusters, with the number of stars for the cluster listed.}
    \begin{tabular}{ccrrrcc} 
    \hline
    Cluster & Age\,(Gyr) & [Fe/H]$_\mathrm{mean}$ & $\ap_\mathrm{mean}$ & [P/Fe]$_\mathrm{mean}$ & $N$ & Alt Name \\
    \hline
    Gulliver 18 & $0.04$ & $-0.01\pm0.02$ & $5.59\pm0.11$ & $0.25\pm0.11$ & 1 & Collinder 416 \\
    Collinder 463 & $0.11$ & $-0.06\pm0.01$ & $5.46\pm0.19$ & $0.16\pm0.19$ & 2 &  \\
    UPK 219 & $0.15$ & $0.07\pm0.01$ & $5.33\pm0.06$ & $-0.10\pm0.06$ & 1 &  \\
    Tombaugh 5 & $0.19$ & $0.02\pm0.06$ & $5.31\pm0.14$ & $-0.08\pm0.15$ & 3 &  \\
    NGC 7086 & $0.19$ & $-0.08\pm0.03$ & $5.61\pm0.17$ & $0.33\pm0.17$ & 2 & Collinder 437 \\
    UBC 194 & $0.23$ & $0.06\pm0.01$ & $5.44\pm0.12$ & $0.02\pm0.13$ & 1 &  \\
    Basel 11b & $0.23$ & $0.03\pm0.03$ & $5.41\pm0.05$ & $0.02\pm0.06$ & 3 & FSR 877 \\
    NGC 2437 & $0.30$ & $0.01\pm0.06$ & $5.45\pm0.28$ & $0.08\pm0.29$ & 4 & M 46, Melotte75 \\
    UBC 169 & $0.30$ & $0.05\pm0.03$ & $5.40\pm0.15$ & $-0.01\pm0.16$ & 2 &  \\
    NGC 2548 & $0.40$ & $0.05\pm0.05$ & $5.30\pm0.07$ & $-0.11\pm0.09$ & 3 & M 48, Melotte 85 \\
    Stock 2 & $0.40$ & $-0.00\pm0.04$ & $5.30\pm0.12$ & $-0.06\pm0.12$ & 7 &  \\
    NGC 7209 & $0.43$ & $0.04\pm0.01$ & $5.63\pm0.05$ & $0.22\pm0.05$ & 1 & Melotte 238, Collinder 444 \\
    Collinder 350 & $0.59$ & $0.08\pm0.01$ & $5.30\pm0.04$ & $-0.15\pm0.04$ & 1 &  \\
    NGC 2632 & $0.68$ & $0.13\pm0.01$ & $5.62\pm0.18$ & $0.13\pm0.19$ & 1 & M 44, Praesepe \\
    NGC 752 & $1.17$ & $0.08\pm0.02$ & $5.30\pm0.08$ & $-0.13\pm0.09$ & 3 & Melotte 12, Theia 1214 \\
    IC 4756 & $1.29$ & $-0.00\pm0.04$ & $5.39\pm0.05$ & $0.03\pm0.06$ & 6 & Collinder386, Melotte 210 \\
    Alessi 1 & $1.45$ & $-0.01\pm0.01$ & $5.49\pm0.01$ & $0.14\pm0.02$ & 4 & Casado-Alessi 1 \\
    NGC 6991 & $1.55$ & $0.03\pm0.06$ & $5.42\pm0.15$ & $0.03\pm0.16$ & 5 &  \\
    UBC 141 & $2.09$ & $-0.02\pm0.01$ & $5.38\pm0.23$ & $0.05\pm0.23$ & 1 &  \\
    UBC 577 & $2.75$ & $-0.04\pm0.04$ & $5.41\pm0.12$ & $0.09\pm0.13$ & 3 & Alessi 191 \\
    Ruprecht 171 & $2.75$ & $-0.04\pm0.05$ & $5.71\pm0.08$ & $0.39\pm0.09$ & 4 &  \\
    NGC 2682 & $4.27$ & $0.02\pm0.01$ & $5.66\pm0.53$ & $0.28\pm0.53$ & 1 & M67 \\
    \hline
    \end{tabular}
    \label{tab:p-cluster}
\end{table*}

\begin{figure*}
    \centering
    \includegraphics[width=1\linewidth]{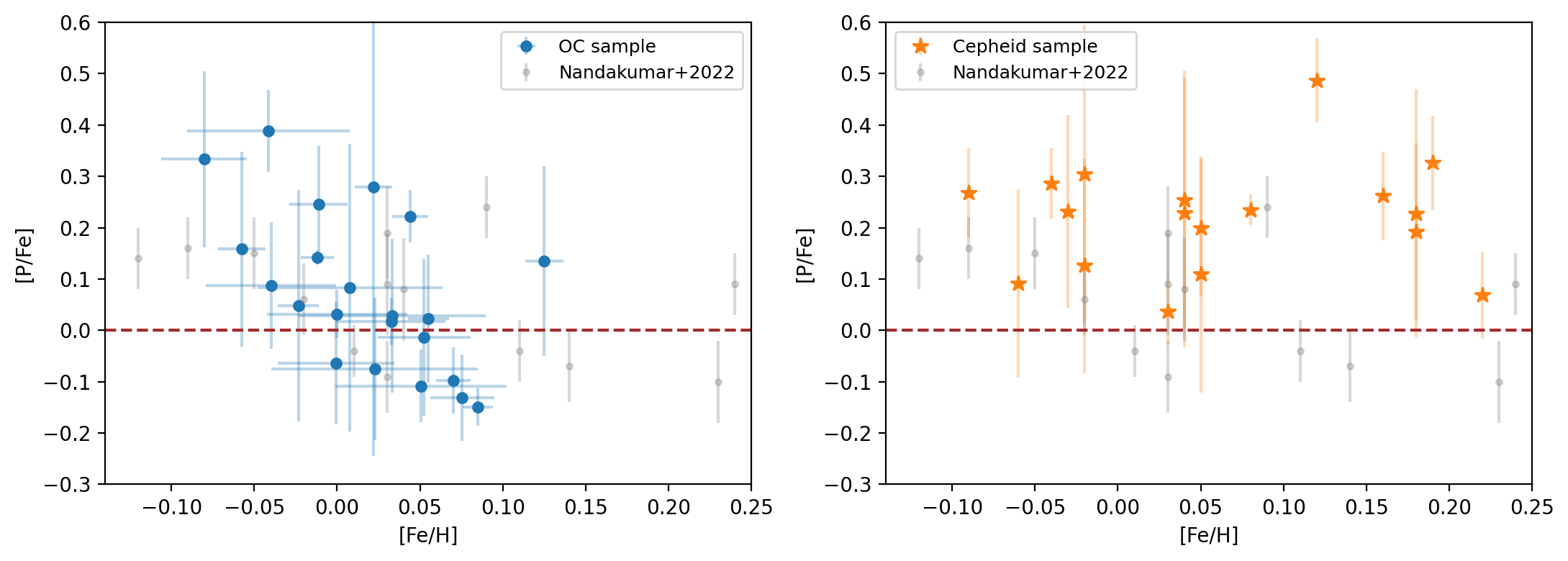}
    \caption{[P/Fe] versus [Fe/H] for OC giants in our sample (left panel, the average $\pfe$ for each cluster) and Cepheid stars (right panel). For comparison, the $\pfe$ measurement of the giants from \citet{Nandakumar2022} is also plotted.
    }
    \label{fig:FeH-AP}
\end{figure*}

The $\pfe$--$\feh$ distribution of our cluster and Cepheid sample is broadly consistent with previous studies, as shown in Figure~\ref{fig:FeH-AP}.
Our targets span a relatively narrow metallicity range, from $\feh = -0.1$ to $+0.15$.
Given this limited coverage, we adopt the results from \citet{Nandakumar2022} as the primary comparison, and refer the reader to their Figure~4 for a comprehensive compilation of earlier studies.
Both our field Cepheids and cluster stars exhibit a modest decreasing trend in $\pfe$ from $\sim 0.3$ to $\sim 0$ over the range $\feh = -0.1$ to $0$.
At the metal-rich end, $\pfe$ appears to flatten near zero, although this trend is not well constrained due to the small number of stars in this regime.
Overall, our observations fall within the same $\pfe$ range as reported by \citet{Nandakumar2022}, supporting the consistency of our results with previous measurements.

\begin{figure}
    \centering
    \includegraphics[width=1\linewidth]{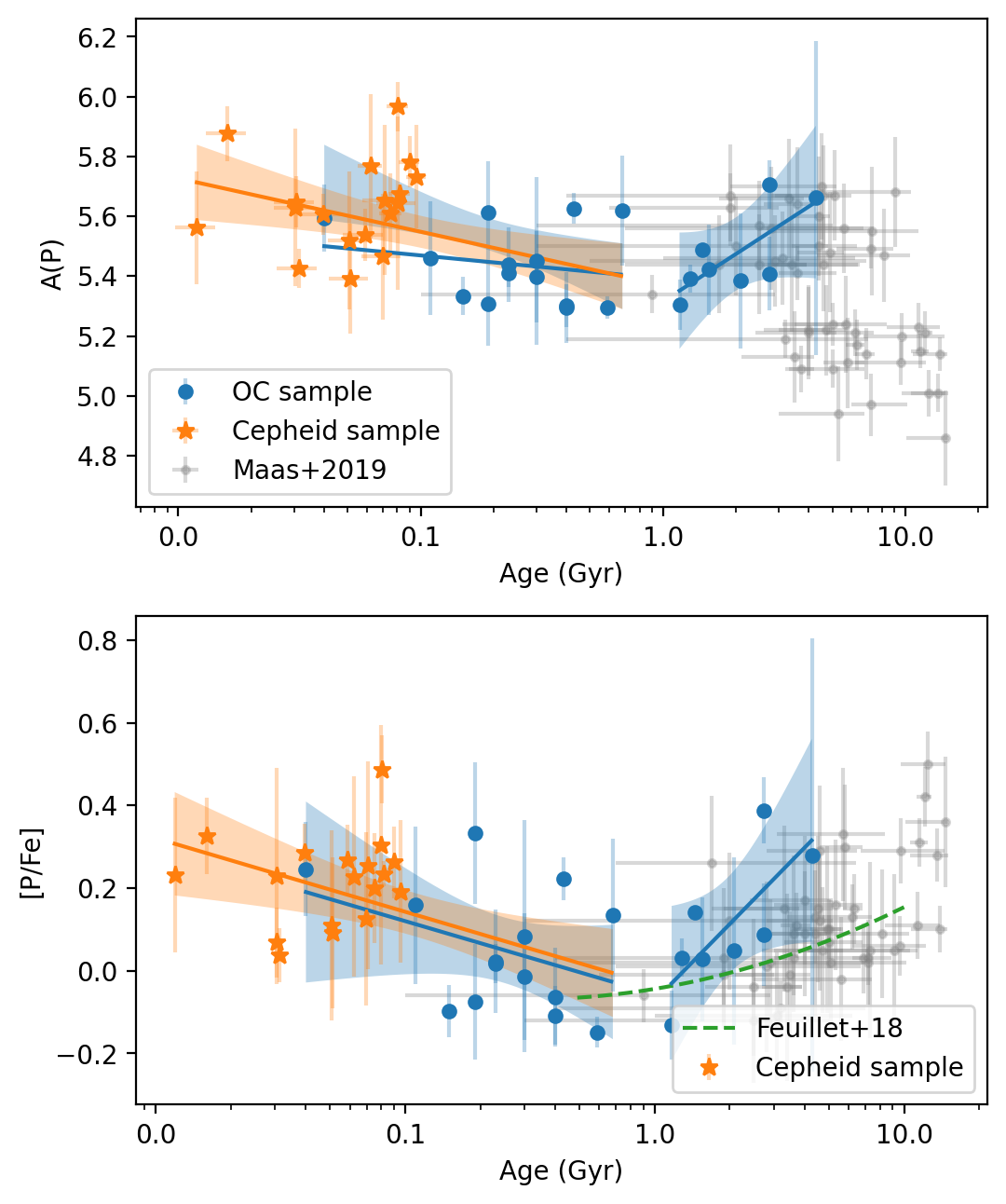}
    \caption{Cluster or stellar age versus phosphorus abundance for our targets. The $\mathrm{\ap}$ and $\pfe$ measurement from \citet{Maas2019} are plotted in gray as a reference. The blue lines and transparent strip represent separate linear fits and its 95\% confidential interval to the cluster data, applied independently to those younger and older than 1\,Gyr, while the orange ones are the linear fit to the cluster and Cepheid sample in the young side. The $\pfe$--age trend from \citet{Feuillet2018} is plotted as a reference.}
    \label{fig:age-AP}
\end{figure}

The $\pfe$--age relation observed in our cluster sample shows general consistency with previous studies, while also providing new insights at the younger end.
As shown in Figure~\ref{fig:age-AP}, the phosphorus abundance (both in the form of absolute P abundance $\ap$ and the relative $\pfe$) of clusters with ages greater than 1\,Gyr increases with age.
This trend is consistent with the findings of \citet{Maas2019} and \citet{Feuillet2018}, although we note that the stellar ages from \citet{Maas2019} (with mean error of $\sim 1.8\,$Gyr) were derived from isochrone fitting of field stars, a method notoriously subject to larger uncertainties than when it is applied to stellar clusters (with relative error around 25\%; \citealt{Cantat-Gaudin2020}).
In contrast, the trend at younger ages seems to be flat.
We performed linear fit to the clusters with ages younger and older than $1\,$Gyr separately in Figure~\ref{fig:age-AP}. 
The linear fit for the older clusters show a positive slope and significant Pearson $p$-values less than $0.05$, in both $\ap$ and $\pfe$.
Those for the young clusters present a negative slope with $p$-values larger than 0.05.

Most clusters with their ages less than $1$\,Gyr have near-solar phosphorus abundances, while the clusters and Cepheids with even younger ages show either solar or enhanced $\pfe$ values.
They reduce the slope of linear fitting and further decrease the $p$-values to less than 0.05.
The measurement of the Cepheids fill the blank in the youngest side of the trend.
They extends the negative trend to the age of $0.015\,$Gyrs, and raises the fitted Phosphorus abundance to $\sim 5.7$.

\subsection{Spatial distribution of Phosphorus}

One of the goal of Galactic archaeology is to map the spatial distribution of elements in our Galaxy. Since OCs represents a well-defined single stellar population, their spatial distribution allows us to investigate the phosphorus distribution among different groups of stars in various directions across the Galactic plane. Classical Cepheids in our sample can also provide similar information, as they are relatively young and their period–luminosity relation enables precise distance measurements.

\begin{figure*}
    \centering
    \includegraphics[width=0.95\linewidth]{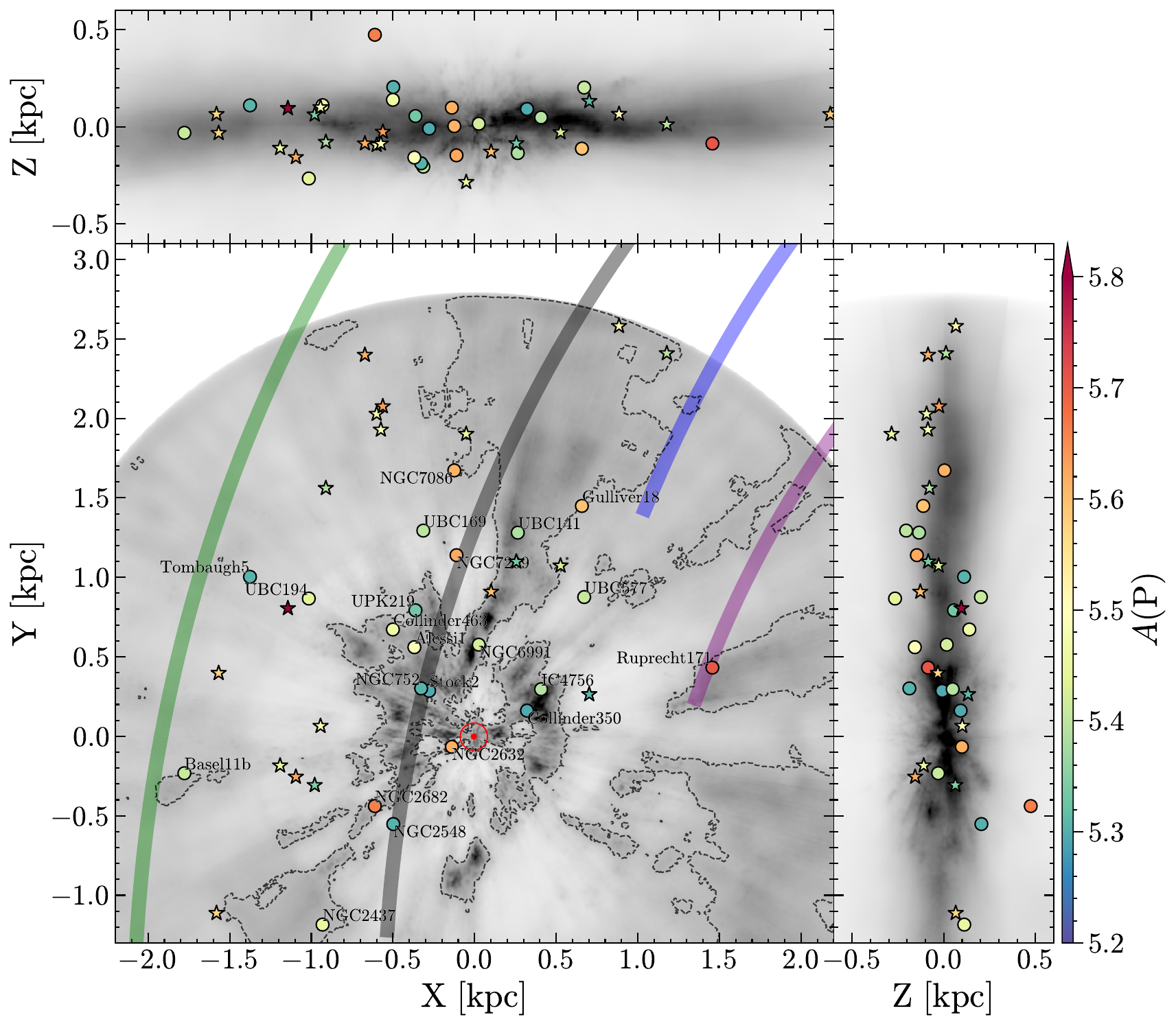}
    \caption{Space distribution of the target clusters (dots) and Cepheids (stars), colour-coded by their phosphorus abundance. 
    The overlaid extinction maps for each plane are derived from \citet{Dharmawardena2024}. 
    A maximum $A_0$ of $0.8$\,mag was set (represented by darkest black), with a contour (black dashed line) at $0.2$\,mag for the XY plane, while a maximum $A_0$ of 3\,mag was applied for the XZ and YZ planes. 
    The coloured curves represent the Galactic log-periodic spiral arms, defined by the parameters from \citep{Reid2019}: the Carina–Sagittarius arm in purple, the Local arm in black, and the Perseus arm in green. 
    Additionally, the spur between the Local and Sagittarius–Carina arms is indicated by the blue curve.}
    \label{fig:P-spatial-distribution}
\end{figure*}

Figure~\ref{fig:P-spatial-distribution} displays the spatial distribution of our target clusters and Cepheids in heliocentric Cartesian coordinates, overlaid with a dust extinction map \citep{Dharmawardena2024} to represent the current structure of the interstellar medium (ISM). 
We adopted Solar $X$ and $Z$ values of $8122$ and $20.8$\,pc, respectively \citep{GRAVITY_Collaboration, Bennett2019}.

Despite the clear inhomogeneous in the present-day ISM, we find no statistically significant correlation between the absolute phosphorus abundance, A(P), and the clusters' current locations, be it their Galactocentric radius or their position relative to local dust structures. This apparent lack of correlation is not a deficiency in our data but is, in fact, a key piece of evidence for the dynamic nature of the Milky Way's disk. Over timescales of tens of millions to several billion years, processes such as radial migration  and vertical heating act to redistribute stars and clusters from their birthplaces \citep[see e.g.][]{VV2023}. 
Consequently, a cluster's current location is often decoupled from the chemical composition of the gas from which it formed.

\section{Discussion}
\label{sec:discussion}

\begin{figure*}
    \centering
    \includegraphics[width=0.9\linewidth]{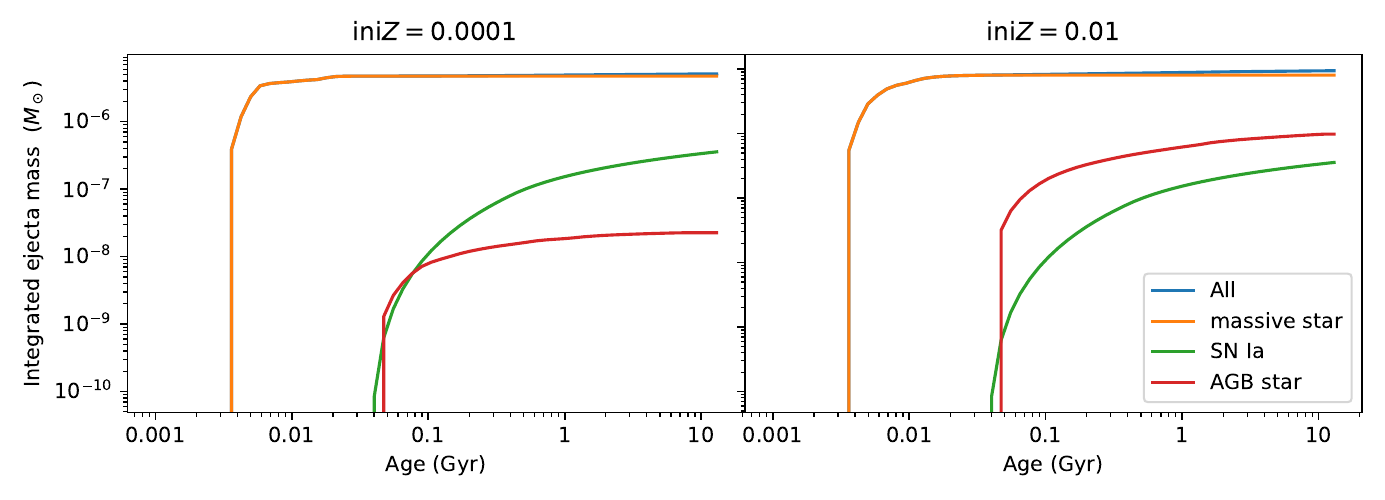}
    \caption{
    Integrated ejecta mass of a single stellar population, with the initial metallicity $\mathrm{ini}Z$ of 0.0001 in the left panel and 0.01 in the right panel.
    }
    \label{fig:P-mass}
\end{figure*}

Our analysis of OCs ages reveals a  distinct ``V-shape'' in the phosphorus enrichment history of the Milky Way's thin disk (see Figure~\ref{fig:age-AP}). While previous studies have established a general trend of $\pfe$ with $\feh$ \citep{Caffau2011, Caffau2016, Maas2019, Maas2022}, the high-precision and wide range of ages with a similar $\feh$  of our cluster sample allow us to dissect the phosphorus evolution in time, uncovering two distinct evolutionary trends.
The ``V-shape'' in the $\ap$-age plane is not a single evolutionary track, but rather the superposition of two different regimes of GCE. 
The clusters younger than $\sim 1\,$Gyr appear to follow a path consistent with local, recent enrichment, while the older clusters unveil a fossil record of the diverse and dynamic star formation histories from the Galaxy's past.
We note that similar but milder $\mathrm{[Mg/Fe]}$--age and $\mathrm{[Mg/H]}$--age trend can also be found using the Mg abundances from \citet{DalPonte2025} where only open clusters are analysed.
A systematic comparison of age--abundance trends for other elements with similar nucleosynthetic origins would be highly informative; however, such an analysis is beyond the scope of the present study and is deferred to future work.

\subsection{Old OCs as a fossil record of diverse star formation histories and P enrichment}

For OCs older than $\sim$1 Gyr, we observe that the absolute phosphorus abundance $\ap$ decreases as stellar age decreases from $\sim$4\,Gyr to $\sim$1\,Gyr. This trend is contrary to the predictions of a simple, single-zone GCE model where metal abundance should monotonically increase over the evolution time. The explanation lies in the fact that the Milky Way is not a single, uniform entity. The solar neighbourhood today contains a mix of stellar populations born in different locations and at different times, each with a unique star formation history. This is  a key feature of multi-zone GCE. 

The clusters in our sample, despite having a similar near-Solar metallicity (probed by $\feh$), reached this metallicity at vastly different cosmic epochs (see Table~\ref{tab:p-cluster}). For a cluster like M67 which has achieved the solar metallicity by about 4\,Gyr ago, its parent gas cloud must have undergone a very rapid and intense period of enrichment. 
In contrast, for another cluster in the old OC group,  NGC 752, its parent gas cloud reached a similar metallicity much more gradually, about 1\,Gyr ago. This difference in enrichment speed implies a fundamentally different Star Formation History.
The rapid enrichment required for the older clusters necessitates either a significantly higher Star Formation Rate or a more ``top-heavy'' IMF that produces more massive stars per generation. 
In either scenario, the integrated contribution from massive stars—the primary producers of phosphorus—is far greater (see Figure~\ref{fig:P-yield}). 
Therefore, it is an inevitable consequence that the 4\,Gyr old M67 formed from gas that was more abundant in absolute phosphorus $\ap$ than the gas that formed the 1\,Gyr-old NGC 752. 
The declining $\ap$ trend for clusters older than $\sim$1\,Gyr is thus a direct manifestation of observing a sequence of clusters born from environments with progressively less intense star formation histories.

\subsection{Young OCs and Cepheids as evidence for local, quiescent P Enrichments}

Open clusters younger than $\sim$1 Gyr and Cepheids in Figure~\ref{fig:age-AP} exhibit a flat or slightly increasing $\ap$ (or $\pfe$) with decreasing age. 
This behaviour aligns well with the expectations for a single-zone GCE model. 
These clusters and stars are young and located in the Solar neighbourhood, meaning they were likely born locally and share a nearly common, recent chemical history. 
They represent the current, more quiescent phase of evolution in our part of the Galaxy. 
In this local environment, we see a slow enrichment of phosphorus over the last $\sim 1\,$Gyr. 

Such gentle phosphorus enrichment may be explained by the contribution from low-mass stars.
Figure~\ref{fig:P-mass} shows the evolution of phosphorus in single stellar population at two different initial metallicity ($\iniZ$) values, following the same set-up as the yields in Figure~\ref{fig:P-yield}. 
Overall, the main phosphorus production at both lower and higher metallicities originates from massive stars, which is consistent with the conclusions of previous studies \citep[e.g.][]{Cescutti2012}. 
The contribution from lower mass stars, i.e. AGB stars, reached a level of $10^{-8}\, M_\odot$ in the later evolution time at $\iniZ = 0.0001$. 
However, the contribution of AGB stars to the phosphorus production at $\iniZ = 0.01$ is an order of magnitude higher than in the metal-poor case, making it more significant at later times in the evolution of the single stellar population. 
This increased phosphorus production by AGB stars is reflected in the peak of the IMF-weighted total yield at the lowest stellar masses, as seen for $\iniZ = 0.004$ and $0.01$ in Figure~\ref{fig:P-yield}. 
These slight sub-Solar metallicities are also proximally corresponding to the previous generation of the OCs and Cepheids in our study.
In the relative absence of vigorous massive star formation in the immediate Solar vicinity, the enrichment from these lower-mass stars can become a noticeable contributor to the ongoing chemical evolution, explaining the slight upward trend in $\ap$ towards the present day.

We note that this interpretation is sensitive to the uncertainties in the nucleosynthetic yields adopted here, highlighting the need for more precise yield determinations.
Also the contribution of ONe nova, recently discussed by \citet{Bekki2024}, is not considered here, but it may contribute to some extent to phosphorus production.

\section{Summary}
\label{sec:conclusion}

In this work, we present a detailed abundance analysis of phosphorus for 82 giant stars in 24 open clusters, together with 20 Cepheids (17 in the Galactic field and 3 in open clusters). 
The analysis is based on high-resolution near-infrared spectra obtained with the GIANO-B spectrograph. 
To achieve this, we developed and implemented a robust line selection and fitting procedure, which carefully accounts for blending, telluric contamination, and line-strength variations, ensuring reliable abundance measurements.

Our key findings include:
\begin{itemize}
    \item Our results confirm the previously observed modest decreasing trend of $\pfe$ with increasing $\feh$ for stars around solar metallicity.
    \item Using the high precision age of open clusters, we uncover a distinct ``V-shape'' in the phosphorus-age relation, revealing two separate chemical evolution pathways and mixtures for the Milky Way's thin disk: \\
    \begin{itemize}
        \item For clusters older than $\sim$1 Gyr in our sample, we find that phosphorus abundance increases with stellar age. We interpret this trend as a fossil record of diverse star formation histories and phosphorus enrichments. With a similar metallicity, older clusters formed from a gas that was more rapidly enriched, a consequence of more intense star formation in the earlier epochs of the Galactic disk. \\
        \item For clusters younger than $\sim$1\,Gyr and Cepheids in our sample, the $\pfe$ and $\ap$ trend with age is nearly flat with a small negative slope. This points to a more quiescent, local enrichment history, where the gentle increase in phosphorus of the younger clusters may be attributable to the contribution from low-mass stars at solar metallicities.
    \end{itemize}
    \item We find no correlation between the $\ap$ abundance and the current spatial location of open clusters and cepheids. 
\end{itemize}

Future observations of more young open cluster giant stars or Cepheids may further clarify the $\pfe$--age trend in the very young population of our Milky Way.

\section*{Acknowledgements}
A.B., M.J., and V.D. acknowledge funding from INAF Mini-Grant 2022 (High resolution spectroscopy of open clusters).
X.F. acknowledges the support of the National Natural Science Foundation of China (NSFC) No. 12203100.
G.B., G.F. and A.N. thank the support from Project PRIN MUR 2022 (code 2022ARWP9C) `Early Formation and Evolution of Bulge and HalO (EFEBHO)' (PI: M. Marconi), funded by the European Union—Next Generation EU, and from the Large grant INAF 2023 MOVIE (PI: M. Marconi). 
Part of the research activities described in this paper were carried out with the contribution of NextGenerationEU funds within the National Recovery and Resilience Plan (PNRR), Mission 4–Education and Research, Component 2–From Research to Business (M4C2), Investment Line 3.1–Strengthening and creation of Research Infrastructures, Project IR0000034— “STILES–Strengthening the Italian Leadership in ELT and SKA”, CUP C33C22000640006.
This work has made use of the VALD database, operated at Uppsala University, the Institute of Astronomy RAS in Moscow, and the University of Vienna.

\section*{Data Availability}

Full version of Figures~\ref{fig:P-line-fit-exmaple-sun}, \ref{fig:P-line-fit-exmaple-Alessi_1_2} and \ref{fig:P-fit_example} are available in the Supporting Information, and Tables~\ref{tab:obslog}, \ref{tab:stellar-paras}--\ref{tab:p-lines-cepheid}, \ref{tab:C-blend-giant} and \ref{tab:C-blend-cepheid} are available at the CDS via anonymous ftp to \url{cdsarc.cds.unistra.fr} (\url{130.79.128.5}), or via \url{https://cdsarc.cds.unistra.fr}.
The data underlying this article will be shared on reasonable request to the corresponding author.



\bibliographystyle{mnras}
\bibliography{refs} 




\appendix

\section{Carbon abundances from blending molecular features}
\label{app:C-blend}

This section presents Table~\ref{tab:C-blend-giant} and \ref{tab:C-blend-cepheid} of C abundance used to fit the molecular features around the phosphorus lines, for cluster giants and Cepheids respectively.

\begin{table*}
    \centering
    \caption{Carbon abundance used to fit molecular features around the phosphorus lines for cluster giants. Only the first 10 stars and first 6 lines are listed, and the full table is available in CDS.}
    \begin{tabular}{clllllll} 
    \hline
    star name & $\ac_{9750.748}$ & $\ac_{10511.588}$ & $\ac_{10529.524}$ & $\ac_{10581.577}$ & $\ac_{10596.903}$ & $\ac_{11183.24}$ \\
    \hline
    Alessi 1-2 & $8.30\pm0.05$ & $8.14\pm0.38$ & $8.56\pm0.10$ & $8.33\pm0.24$ & $8.06\pm0.22$ & $8.31\pm0.08$ \\
    Alessi 1-3 & $8.27\pm0.06$ & $8.25\pm0.28$ & $8.64\pm0.09$ & $8.40\pm0.22$ & $8.42\pm0.08$ & $8.25\pm0.08$ \\
    Alessi 1-5 & $8.38\pm0.05$ & $8.04\pm0.37$ & $8.61\pm0.08$ & - & $8.40\pm0.06$ & $8.30\pm0.07$ \\
    Alessi 1-6 & $8.26\pm0.07$ & $7.68\pm0.78$ & $8.71\pm0.08$ & - & $7.63\pm0.16$ & $8.20\pm0.09$ \\
    Alessi Teutsch 11-1 & $8.25\pm0.05$ & $8.35\pm0.11$ & $8.52\pm0.05$ & - & $8.32\pm0.03$ & - \\
    Basel 11b-1 & $8.35\pm0.07$ & $8.47\pm0.14$ & $8.79\pm0.07$ & - & $8.15\pm0.25$ & $8.33\pm0.09$ \\
    Basel 11b-2 & $8.42\pm0.06$ & $8.37\pm0.19$ & $8.79\pm0.07$ & $8.67\pm0.12$ & $8.51\pm0.08$ & $8.38\pm0.08$ \\
    Basel 11b-3 & - & - & - & $8.54\pm0.09$ & - & - \\
    COIN-Gaia 30-1 & $7.87\pm0.17$ & $8.12\pm0.29$ & $8.52\pm0.09$ & - & $8.38\pm0.07$ & $8.45\pm0.05$ \\
    Collinder 350-1 & - & - & - & - & - & - \\
    Collinder 350-2 & - & - & - & $8.55\pm0.14$ & - & - \\
    ...  &...&...&...&...            &... &... \\
    \hline
    \end{tabular}
    \label{tab:C-blend-giant}
\end{table*}

\begin{table*}
    \centering
    \caption{Carbon abundance used by fitting the molecular features around the phosphorus lines for Cepheids. Only the first 10 stars and first 6 lines are listed, and the full table is available in CDS.}
    \begin{tabular}{cllllll} 
    \hline
    star name & $\ac_{9493.571}$ & $\ac_{9609.036}$ & $\ac_{10932.724}$ & $\ac_{10967.373}$ & $\ac_{11186.753}$ & $\ac_{15711.522}$ \\
    \hline
    SV Vul & $8.08\pm0.18$ & $8.64\pm0.04$ & $8.52\pm0.11$ & $8.63\pm0.02$ & $8.41\pm0.10$ & $10.27\pm0.09$ \\
    DL Cas & - & - & - & - & - & - \\
    CO Aur & $6.35\pm2.36$ & - & - & $8.35\pm3.05$ & - & - \\
    RX Aur & - & - & - & - & - & - \\
    V 351 Cep & - & - & - & $8.33\pm2.73$ & $10.33\pm0.31$ & - \\
    CD Cyg & - & - & $8.25\pm0.02$ & - & - & - \\
    V 1334 Cyg & $6.09\pm5.05$ & - & - & - & - & - \\
    VZ Cyg & $8.07\pm20.49$ & $10.06\pm0.61$ & - & $8.07\pm4.22$ & - & - \\
    X Cyg & - & - & - & - & - & - \\
    W Gem & - & - & - & - & - & - \\
    RR Lac & - & - & - & - & - & - \\
    ...  &...&...&...&...            &... &... \\
    \hline
    \end{tabular}
    \label{tab:C-blend-cepheid}
\end{table*}

\section{The P lines with biased abundance}
\label{app:unused-P}

As described in Section~\ref{sec:abund_measure} and summarised in Table~\ref{tab:p_lines_all}, four lines selected for cluster giants and four lines for Cepheids were excluded from the analysis due to systematic biases in their derived phosphorus abundances.
Among these, two lines for the giants and one for the Cepheids were used in more than five stars.
Here, we discuss these lines individually.

The \ion{P}{i} line at 10813.141\,\AA{} is intrinsically very weak, with a typical line depth of only $\sim$0.01 in most stars.
Its abundance sensitivity is highly dependent on accurate continuum placement.
However, the continuum in the fitting region is often depressed below unity, and since \texttt{PySME} applies only a constant scaling over a wide region during abundance fitting, local misplacement of the continuum across a narrow wavelength range cannot be corrected.
As a result, the phosphorus abundances derived from this line tend to be overestimated.
Also, the situation for \ion{P}{i} line at 17423.670\,\AA{} is similar but with underestimated abundance, thus it is also being removed.

The line at 17112.447\,\AA{} consistently yields phosphorus abundances that are approximately 0.3\,dex higher than the average derived from other lines.
Inspection of the line fitting reveals the presence of an unidentified absorption feature on the red side of the P line.
This feature is absent from the synthetic spectra (otherwise it would have been accounted for), but it appears consistently in all observed spectra with a similar relative position, suggesting it originates from the stellar spectrum.
Due to the severity of this blending, the fitting routine attempts to reproduce the composite feature by increasing the phosphorus abundance, leading to a systematic overestimation.

Six \ion{P}{i} lines were excluded from the Cepheid sample. Their behavior is similar to that of the 10813.141\,\AA{} line, for which accurate continuum placement is challenging in this wavelength region, compromising the reliability of the abundance fitting. As a result, these lines were excluded from the analysis.




\bsp	
\label{lastpage}
\end{document}